# SACRE: Supporting contextual requirements' adaptation in modern self-adaptive systems in the presence of uncertainty at runtime

Edith Zavala[*], Xavier Franch[*], Jordi Marco[*], Alessia Knauss[†], Daniela Damian[±]

zavala@essi.upc.edu, franch@essi.upc.edu, jmarco@cs.upc.edu,

alessia.knauss@chalmers.se, danielad@cs.uvic.ca

[*]*Software and Service Engineering research group (GESSI), Universitat Politècnica de Catalunya (UPC), Jordi Girona, 1-3, 08034, Barcelona, Catalunya, Spain*

[†]*Department of Computer Science and Engineering, Chalmers and the University of Gothenburg, Hörselgången 5, 41296 Göteborg, Sweden*

[±]*Department of Computer Science, University of Victoria, 3800 Finnerty Rd, Victoria, BC V8P 5C2, Canada*

**Corresponding author:**

Edith Zavala

Software and Service Engineering research group (GESSI)

Universitat Politècnica de Catalunya (UPC)

Jordi Girona, 1-3, 08034

Barcelona, Catalunya, Spain

Tel: +34 644 20 22 23

Email: zavala@essi.upc.edu

## ABSTRACT

Runtime uncertainty such as unpredictable resource unavailability, changing environmental conditions and user needs, as well as system intrusions or faults represents one of the main current challenges of self-adaptive systems. Moreover, today's systems are increasingly more complex, distributed, decentralized, etc. and therefore have to reason about and cope with more and more unpredictable events. Approaches to deal with such changing requirements in complex today's systems are still missing. This work presents SACRE (Smart Adaptation through Contextual REquirements), our approach leveraging an adaptation feedback loop to detect self-adaptive systems' contextual requirements affected by uncertainty and to integrate machine learning techniques to determine the best operationalization of context based on sensed data at runtime. SACRE is a step forward of our former approach ACon which focus had been on adapting the context in contextual requirements, as well as their basic implementation. SACRE primarily focuses on architectural decisions, addressing self-adaptive systems' engineering challenges. Furthering the work on ACon, in this paper, we perform an evaluation of the entire approach in different uncertainty scenarios in real-time in the extremely demanding domain of smart vehicles. The real-time evaluation is conducted in a simulated environment in which the smart vehicle is implemented through software components. The evaluation results provide empirical evidence about the applicability of SACRE in real and complex software system domains.

## 1. Introduction

Nowadays, software systems are extremely capable, mimicking natural systems' characteristics such as intelligence, rationality, ability to learn, anticipation, and automatic adaptation to context. These kind of systems are called self-adaptive systems (SASs) (De Lemos et al., 2013; Krupitzer et al., 2015). Concretely, a SAS is a system able to automatically modify itself in order to respond to changes in its environment and the system itself (Cheng et al., 2009). A particular type of requirements, contextual requirements, has been proposed for supporting the adaptation capabilities in SASs and keeping track of user needs and environmental conditions at the same time (Ali, 2010; Bhaskar & Govindarajulu, 2008; Inverardi & Mori, 2013; Knauss et al., 2016) . Based on previous work (Knauss et al., 2016): A *contextual requirement* consists of a 2-tuple of the <<*expected system behavior*>> and the specific <<*context*>> within which this expected behavior is valid. For instance, in the domain of smart vehicles, an example of contextual requirement is: when it is raining (<<*context*>>), the system shall activate the windshield wiper (<<*expected system behavior*>>).

Due to their intensive interaction with environments, SASs are inevitably affected by runtime uncertainty. *Uncertainty* is a system state of incomplete or inconsistent knowledge such that it is not possible for the SAS to know which of two or more alternative environmental or system configurations hold at a specific point (Ramirez et al., 2012). As SASs grow in complexity, they must increasingly reason about and cope with unpredictable environments (Ramirez et al., 2012). Thus, managing uncertainty is particularly challenging in modern and complex software application domains like smart cities, smart vehicles and mobile apps, in which intensive user interaction, heterogeneity, pervasiveness, varying decentralization levels and distribution are very important factors (Cheng et al., 2009; De Lemos et al., 2013; Krupitzer et al., 2015; Weyns, 2017). These systems must find a way to deal with uncertainty, handling complex issues as unpredictable resource unavailability, changing environmental conditions and user needs, system intrusions or faults, and sensors' decalibration. These issues imply systems' ability to deal with 1) emerging requirements not present at design time and 2) requirements that change or even disappear during system operation, with the purpose of improving user satisfaction and system's effectiveness, availability, and robustness.

From the state-of-the-art analyzed in this work, we have identified only a few studies focusing on the automatic support of changing SASs' contextual requirements at runtime, particularly in the presence of uncertainty (Gerostathopoulos et al., 2016; Han et al., 2016; Klos et al., 2015; Knauss et al., 2016); while, to the best of our knowledge, there are no proposals explicitly focusing on addressing this challenge in modern SASs (i.e., pervasive, heterogeneous, distributed and decentralized). In our previous work (Knauss et al., 2016), we developed a machine learning-based approach, ACon, that uses a MAPE-K feedback control loop reference model (IBM-Corporation, 2006) for adapting SASs' contextual requirements affected by uncertainty at runtime. The evaluation results were promising and pointed out the ability of ACon for dealing with the challenge of managing runtime uncertainty in SASs (Knauss et al., 2016). However, ACon faced some limitations regarding the current challenges affecting the operation of modern SASs at runtime (e.g., the need of control decentralization). Moreover, it did not provide details about how the solution should be constructed in actual SASs (e.g., how MAPE-K elements are mapped onto software components) and its evaluation was focused on the performance of the data mining algorithm, thus only part of the loop was implemented.

In this paper we present SACRE, an approach that extends the proposal of ACon for applying the learning-based approach in modern SASs. In order to do so, we propose a reference architecture that captures the basic functionalities proposed by ACon for the MAPE-K reference model, as well as the extended functionalities proposed by SACRE. Furthermore, we introduce an architectural pattern for supporting decentralized and collaborative control in SASs. This pattern, combines and extends existing patterns for decentralized control in SASs proposed by Weyns et al. (2013). The design of the pattern is based on 3 layers, similar to the one proposed by Perrouin et al. (2012), for the evolution of the adaptation logic of SASs. SACRE relies on a policy-based operation and the observer-observable pattern (Gamma et al., 1995) for decoupling the functionalities of the elements of the MAPE-K loop.

The evaluation of SACRE is performed in real-time with an example from the extremely demanding domain of smart vehicles. By smart vehicle, we mean a vehicle that is capable of sensing contextual data (e.g., from the driver, the environment, the vehicle itself) and making decisions based on this data (e.g., turning on an alarm, activating self-driving functionality). This real-time evaluation is conducted in a simulated environment in which the smart vehicle is implemented through software components. Although this running example plays an important role in this paper, the main focus is not on the functionality of the smart vehicle, but on demonstrating the validity of SACRE in modern SASs through the implementation of all feedback loop elements, providing the details about the system's construction, as well as the adaptation response time and accuracy results obtained from this evaluation. A preliminary version of the implementation that is provided in this work has been presented as a demo tool in the IEEE RE'15 conference (Zavala et al., 2015) as a proof-of-concept of the ACon approach in the smart vehicles domain.

The main contribution of this paper is our improved approach SACRE for supporting the adaptation of contextual requirements affected by uncertainty in modern SASs in which decentralization, heterogeneity, pervasiveness, among other factors, are very important. The remainder of the paper is organized as follows: Section 2 discusses the open research challenges that affect the operation of current SASs at runtime. Section 3 presents the state-of-the-art regarding the most important open research challenges we identified in Section 2. Section 4 establishes the research goal of this work. Section 5 presents the fundamentals of our SACRE approach. In Section 6, an overview of the application of SACRE in the domain of smart vehicles is provided. Section 7 presents the technical details about the implementation of SACRE in this domain. Then, in Section 8, we report the empirical evidence about the evaluation of SACRE in real-time. Section 9 discusses the results obtained. Finally, Section 10 draws the conclusions and depicts future work.

## 2. Open research challenges in SASs

Over time extensive efforts have been spent in different research fields on the realization of SASs (Weyns, 2017). However, some challenges regarding the capabilities and construction of SASs remain open. In this section, we investigate different works in order to identify these challenges. Particularly, we focus on challenges that affect the operation of SASs at runtime (e.g., challenges about design-time languages for describing SASs will be out of the scope of this paper). In the first and second Software Engineering for SASs research roadmap papers (Cheng et al., 2009; De Lemos et al., 2013), the authors identified a set of challenges that affect current SASs and motivate the need for research in different fields (i.e., runtime RE, SASs engineering, runtime verification and validation (V&V), adaptation assurance, etc.). Besides these challenges, Krupitzer et al. (2015) added two more important ones regarding the capability of supporting runtime proactive adaptation and context adaptation.

In a more recent work, Weyns (2017), speculating on how the field may evolve in the future, presents what he considers the most worth focusing open research challenges confirming some of the previously identified by Cheng et al. (2009), De Lemos et al. (2013) and Krupitzer et al. (2015), and introducing two more new challenges regarding the integration of the system adaptation and evolution processes, and the support of automated runtime system models. Concretely, from these four resources (Cheng et al., 2009; De Lemos et al., 2013; Krupitzer et al., 2015; Weyns, 2017), we have identified the following open research challenges (**Chl**), which we divide into four categories:

- *Category 1. SASs capabilities challenges*
    - **Chl1.1**. Provide self-adaptation capabilities to existing systems (Cheng et al., 2009).

- **Chl1.2**. Perform trade-off analysis between several potential conflicting system goals (Cheng et al., 2009; Krupitzer et al., 2015).
- **Chl1.3**. Support different adaptation mechanisms (e.g., composition, parameter) for leveraging system capabilities (Cheng et al., 2009).
- **Chl1.4**. Support context adaptation (Krupitzer et al., 2015).
- **Chl1.5**. Support proactive adaptation (Krupitzer et al., 2015).
- **Chl1.6**. Integrate system adaptation with evolution for dealing with unanticipated changes (Weyns, 2017).

- *Category 2. SASs engineering challenges*
    - **Chl2.1**. Perform adaptation activities (i.e., monitoring, analysis, decision-making, execution, V&V) without affecting system performance and availability (Cheng et al., 2009; Krupitzer et al., 2015).
    - **Chl2.2**. Communicate, coordinate and share loop elements with other SASs (Cheng et al., 2009; Krupitzer et al., 2015; Weyns, 2017).
    - **Chl2.3**. Support both centralized and decentralized control loops (De Lemos et al., 2013; Krupitzer et al., 2015; Weyns, 2017).
    - **Chl2.4**. Predict the effects of adaptation, e.g., overhead (Cheng et al., 2009).
    - **Chl2.5.** Support runtime use of system models (machine-driven) (Weyns, 2017).

- *Category 3. SASs requirements challenges*
    - **Chl3.1**. Capture self-adaptation capabilities and runtime uncertainty in requirements (Cheng et al., 2009; Krupitzer et al., 2015; Weyns, 2017).
    - **Chl3.2**. Permit requirements and goals monitoring and adaptation at runtime (Cheng et al., 2009; Krupitzer et al., 2015; Weyns, 2017).
    - **Chl3.3**. Enable dynamic traceability from requirements to implementation (Cheng et al., 2009).
    - **Chl3.4**. Consider the user-in-the-loop (Cheng et al., 2009; Krupitzer et al., 2015).
    - **Chl3.5**. Balance requirements adaptation and assurance such that system high-level goals are always met (Cheng et al., 2009).

- *Category 4. SASs runtime assurance challenges*
    - **Chl4.1**. Identify and verify new contexts at runtime for accurately calculating requirements (Cheng et al., 2009).
    - **Chl4.2**. Sense and recover from failures (De Lemos et al., 2013).
    - **Chl4.3**. Integrate V&V activities (e.g., testing, formal verification, adaptation decisions checking, analysis) in the runtime self-adaptation lifecycle (i.e., control loop) (De Lemos et al., 2013; Krupitzer et al., 2015).

From the reported challenges it can be noticed that some engineering and requirements challenges have prevailed over the years, these are: **Chl2.2**, **Chl2.3**, **Chl3.1** and **Chl3.2**. This does not mean that the rest of challenges have been already addressed by existing approaches but we believe that these prevalent challenges have particular importance and they must be on the top of the SASs research community agenda. In the next section, we analyze the state of the art with respect to these four challenges.

## 3. State of the art

In this section, we analyze, characterize and compare how existing proposals address challenges **Chl2.2**, **Chl2.3**, **Chl3.1** and **Chl3.2**. First of all, we provide the details about the protocol we have followed for identifying the proposals. Then, we present our analysis about the state-of-the-art works.

### 3.1. Study selection protocol

In order to identify approaches that explicitly address the challenges of our interest (**Chl2.2**, **Chl2.3**, **Chl3.1** and **Chl3.2**), we have followed the protocol we describe below:

- **Stage 1**. In this first stage, we considered the literature cited by the studies we have used for identifying the open research challenges in Section 2, i.e., Cheng et al. (2009), De Lemos et al. (2013), Krupitzer et al. (2015) and Weyns (2017). From this stage, 364 papers (once duplicates were automatically removed) have been identified, as it is shown in Fig. 1.
- **Stage 2**. From the set of studies identified in Stage 1, the first author has extracted the approaches supporting the engineering of collaborative and/or decentralized SASs control loops (i.e., approaches related to **Chl2.2** and **Chl2.3**) as well as approaches dealing with the adaptation of SASs requirements, particularly in the presence of uncertainty (i.e., approaches related to **Chl3.1** and **Chl3.2**). As we show in Fig. 1, from this stage, 33 papers were identified. The list of references resulting from this stage can be found in Appendix A.

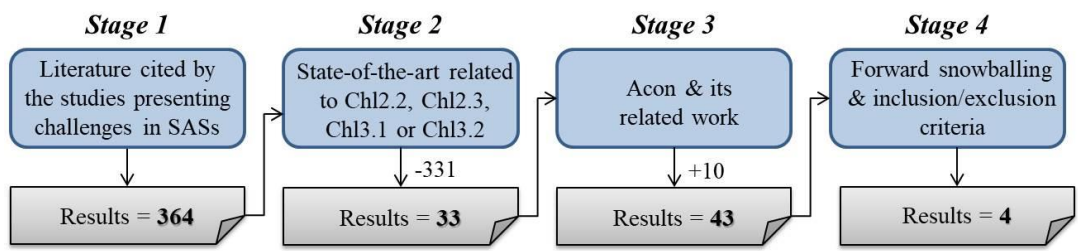

*Fig. 1. Number of included articles during the selection process*

**Stage 3**. In this stage, we considered our former approach Acon (Knauss et al., 2016) and the state-of-the-art approaches related to contextual requirements adaptation in the presence of uncertainty presented in its related work. This provided us 10 more papers (see Fig. 1). The list of references added in this stage can be found in Appendix B.

- **Stage 4**. Finally, the first author has performed a forward snowballing, in order to identify advances added on top of these approaches or new emerging ones, and inclusion/exclusion process. Studies have been excluded based on titles, abstracts, as well as full-text reading. Periodic meetings with the rest of the authors, during the process of selection for discussing the papers, were done. In this stage, we were interested in related work to SASs requirements adaptation in the presence of runtime uncertainty and in approaches that adds up on top of this with integration concepts for modern self-adaptive systems. Thus, the following inclusion criteria were applied:

- Studies present a solution for supporting the adaptation of SASs requirements at runtime.
- Studies present a solution for dealing with runtime uncertainty affecting SASs requirements, through their adaptation.
- Studies present a solution for supporting requirements adaptation in decentralized or distributed SASs.
- Studies present a solution for adapting SASs requirements and support the communication, sharing and/or coordination of SASs control loop elements with other SASs.

The following exclusion criteria were applied:

- Studies presenting summaries of solutions fulfilling the inclusion criteria, i.e., secondary studies.
- Studies not accessible in full-text.

In order to decide when to stop the snowballing process, the saturation criterion has been used as follows: when an article did not fulfill the inclusion criteria (or fulfilled the exclusion criteria) the snowballing process, for that article, stopped and the article was discarded; when the studies referencing an article, that fulfilled the inclusion criteria, did not fulfill them (or fulfilled the exclusion criteria), the process for that article stopped and the article was added to the final set. From this process, four studies (Gerostathopoulos et al., 2016; Klos et al., 2015; Han et al., 2016; Knauss et al., 2016) have resulted, as we show in Fig. 1. These studies are further analyzed in the rest of this section.

### 3.2. Results

Now, we analyze how the four approaches selected in Section 3.1 (Gerostathopoulos et al., 2016; Klos et al., 2015; Han et al., 2016; Knauss et al., 2016), address challenges **Chl2.2**, **Chl2.3**, **Chl3.1** and **Chl3.2**. A summary of the results of this analysis is presented in Table 1.

*Table 1. Comparing how approaches related to SACRE address most important SASs challenges*

| *Work by* | *Chl2.2* | *Chl2.3* | *Chl3.1* | *Chl3.2* |
|---|---|---|---|---|
| Gerostathopoulos et al. (2016) | Interaction with other SASs is not supported (considered). | Decentralization is not supported (discussed). | Adaptation strategies capture the self-adaptation capabilities. | Adaptation strategies monitoring and adaptation at runtime is supported. |
| Klos et al. (2015) | Interaction with other SASs is not supported (considered). | Decentralization is not supported (discussed). | Adaptation rules capture self-adaptation capabilities. Uncertainty at design time is not captured but managed through runtime models mutation for satisfying goals. | Adaptation rules monitoring and adaptation at runtime is supported. |
| Han et al. (2016) | Interaction with other SASs is not supported (considered) | Decentralization is not supported (discussed). | Fuzzy rules capture self-adaptation capabilities and uncertainty. | Fuzzy rules monitoring and adaptation at runtime is supported. |
| Knauss et al. (2016) | Interaction with other SASs is not supported (considered). | Decentralization is not supported (discussed). | Contextual requirements capture self-adaptation capabilities. Uncertainty at design time is not captured but managed through context' re-operationalization at runtime. | Contextual requirements monitoring and adaptation at runtime is supported. |
| **SACRE** | **Inter-intra collaborative MAPE-K loops are supported.** | **Supports both centralized and decentralized loops through knowledge distribution and coordination mechanisms.** | **Contextual requirements capture self-adaptation capabilities. Uncertainty at design time is not captured but managed through context' re-operationalization at runtime.** | **Contextual requirements monitoring and adaptation at runtime is supported.** |

Gerostathopoulos et al. (2016), present a 3-layer architectural solution for increasing the homeostasis of self-adaptive software-intensive cyber-physical systems, i.e., the capacity to maintain an operational state despite runtime uncertainty, by introducing runtime changes to the self-adaptation strategies, i.e., the SASs adaptation capabilities (**Chl3.1** and **Chl3.2**). The architecture is composed of a set of MAPE-K loops that interact between them hierarchically for supporting architectural adaptations of complex SASs and their adaptation capabilities. However, these MAPE-K loops are centralized and details about how they can interact with loops of other SASs (**Chl2.2** and **Chl2.3**) are not provided.

Klos et al. (2015), proposes an extended architecture of the MAPE-K loop for supporting the adaptation of SASs' adaptation capabilities in the form of rules in order to respond to unanticipated environmental changes (**Chl3.1** and **Chl3.2**). The approach utilizes system, environment and global goal models stored in the Knowledge base (K element of the loop) for automatically evaluate the adequacy of current adaptation rules and delete or generate new rules at runtime through the mutation of runtime models. The MAPE-K loop is extended with two new components: Evaluation and Learning. The internal approach proposed by this solution could affect the performance of the normal MAPE-K operation, introducing unnecessary overhead. As described by Krupitzer et al. (2015), external approaches like the hierarchical adopted by Gerostathopoulos et al. (2016), are preferable since they ease the scalability and maintainability of the system. Moreover, the decentralization of the MAPE-K elements and their collaboration with other MAPE-K loops have not being discussed in this work (**Chl2.2** and **Chl2.3**).

Another interesting work dealing with runtime uncertainty in SASs through the automatic adaptation of SASs adaptation capabilities, is the one presented by Han et al. (2016). This work provides learning abilities to the analysis and plan elements of the loop (combined in a Self-learning adapter component) in charge of continuously evaluate the adaptation rules and trigger SASs adaptations as well as adapt the rules that capture the SASs adaptation capabilities, when necessary, in order to handle runtime uncertainty (**Chl3.1** and **Chl3.2**). This component, analyses runtime sensor data and adapts its rules based on learnings obtained from these data. This approach faces similar limitations as the previous one (Klos et al., 2015), its internal approach and the lack of decentralization and collaboration mechanisms condition its applicability in modern SASs (**Chl2.2** and **Chl2.3**).

Finally, our previous approach ACon (Knauss et al., 2016), proposes the adaptation of SASs' contextual requirements for dealing with runtime uncertainty. The proposal relies on an external MAPE-K feedback loop that interacts with SASs for monitoring and adapting their requirements. However, this work does not provide architectural details for constructing such proposal. Moreover, cooperation and decentralization mechanisms were not provided, limiting the demonstration of its value in modern SASs (**Chl2.2** and **Chl2.3**).

## 4. Research goal

According to the results of the state-of-the-art provided in Section 3, there is not any proposal in the literature studied that covers all the four prevalent challenges affecting SASs identified in this work:

- **Chl2.2.** Communicate, coordinate and share loop elements with other SASs.
- **Chl2.3**. Support both centralized and decentralized control loops.
- **Chl3.1.** Capture self-adaptation capabilities and runtime uncertainty in requirements.
- **Chl3.2**. Permit requirements and goals monitoring and adaptation at runtime.

Therefore, the research goal of this paper is formulating a proposal to address these four challenges. In order to do that, we introduce in the next section our improved approach SACRE. In comparison with the found approaches, as we report in Table 1, SACRE proposes a solution for addressing all the four prevalent challenges. First, it relies on decoupled, flexible and collaborative MAKE-K elements for supporting both centralized and decentralized loops as well as the communication and sharing of SASs loops' elements (**Chl2.2**, **Chl2.3**). Then, it adopts the operation proposed by our former approach ACon (Knauss et al., 2016) for the MAPE-K feedback loop elements in order to support the adaptation of SASs contextual requirements in the presence of runtime uncertainty (**Chl3.1** and **Chl3.2**).

## 5. SACRE

In order to formalize the SACRE approach, in this section we first provide an overview of the MAPE-K reference model adopted by SACRE from ACon (Knauss et al., 2016), i.e., we briefly describe the operation of the feedback loop proposed by our former learning-based approach in order to correctly support SASs' requirements adaptation at runtime in the presence of uncertainty. Then, we present an architectural pattern for supporting decentralized and collaborative MAPE-K feedback loops. Finally, we provide the details about SACRE and how it extends the ACon proposal for addressing **Chl3.1** and **Chl3.2,** with the functionalities required for supporting relations proposed by the architectural pattern for tackling **Chl2.2** and **Chl2.3**. A vision about how SACRE elements interact between them and with other systems is provided in the form of a reference architecture.

### 5.1. MAPE-K reference model in SACRE

In Fig. 2, we briefly describe the MAPE-K feedback loop reference model operation in SACRE. As mentioned before in this section, SACRE adopts the MAPE-K functionalities proposed by our former approach ACon (Knauss et al., 2016). Concretely, the MAPE-K loop, because of its four elements: Monitor, Analyze, Plan, and Execute and the Knowledge base, implements the following adaptation process:

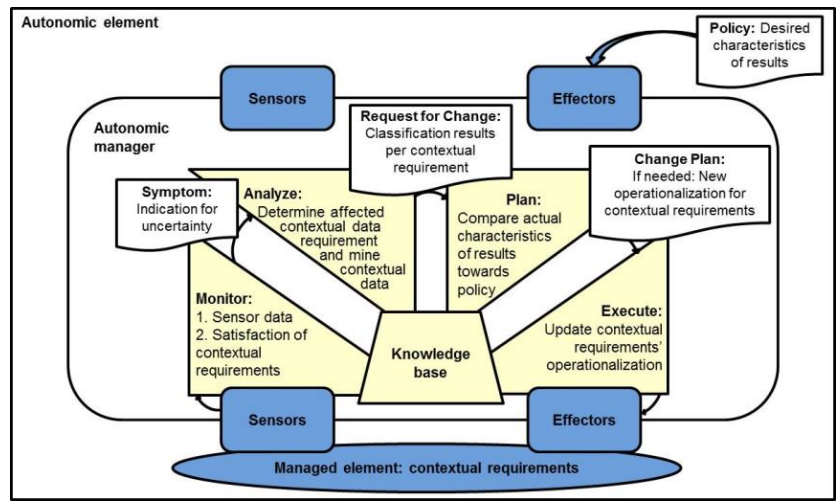

*Fig. 2. Adaptation feedback loop in SACRE adopted from ACon (Knauss et al., 2016)*

- First, the **Monitor** element senses a set of environmental variables and the current set of contextual requirements through a Sensors interface. Then, it determines the state of the requirements' <<*expected system behavior*>> (i.e., active or inactive) and evaluates which contextual requirements' <<*context*>> hold. In order to evaluate contexts, the Monitor uses contexts' operationalization which is expressed in terms of the environmental variables combined by expression operators (i.e., relations, arithmetic, and logical). For example, for the <<*context*>>: when it is raining, its operationalization can be: (humidity>=0,7) AND (cloudy>0,6). With this information, it calculates the satisfaction of contextual requirements, i.e., it corroborates that the states of the requirements' <<*expected system behavior*>> correspond to the <<*context*>> holding at that point. Four runtime uncertainty cases affecting contextual requirements' satisfaction are considered by ACon (see *Table* 2).

*Table 2. Uncertainty cases affecting contextual requirements' satisfaction*

| *Case* | *Detection of uncertainty* |
|---|---|
| **Case 1** | No operationalized context. |
| **Case 2 a)** | Sensor lost. |
| **Case 2 b)** | Sensor decalibrated. |
| **Case 2 c)** | Sensor up (again). |
| **Case 3** | Violation (i.e., requirement's context holds (true) but expected behavior is not active (false)). |
| **Case 4** | Potentially wrong context (i.e., requirement's context does not hold (false) but expected behavior is active (true)). |

When a requirement is dissatisfied (for an arbitrary number of iterations), the Monitor element diagnoses this requirement to be affected by uncertainty and send a Symptom (consisting of the contextual requirement affected and the uncertainty case affecting it) to the Analyze element for further analysis.

- When the **Analyze** element receives a Symptom, it analyzes context data (i.e., environmental and contextual requirements historical data) using a rule-based data mining algorithm for finding patterns and the best operationalization of context for minimizing the number of requirements affected by uncertainty (i.e., dissatisfied). *If…then* rules produced by the data mining algorithm, are easily translated into requirements' context operationalizations. The resulting operationalizations and data mining algorithm statistical measures (e.g., accuracy, error, etc.) as well as the affected analyzed requirement, are sent in a Request for Change to the Plan element.

- Then, the **Plan** element receives the requests and determines the adaptations to execute by considering a given policy (e.g., acceptance thresholds for the resulting data mining algorithm statistical measures) and creates a Change Plan (consisting of the requirement to be adapted as well as the new operationalization) that sends to the Execute element.

- Finally, the **Execute** element uses the Effectors interface for sending the adaptations (i.e., context re-operationalizations) to the SAS.

- All this process is performed sharing a **Knowledge base** that persist relevant information about contextual requirements and environmental data as preparation to apply data mining on.

**5.2. The Hierarchical inter-intra collaborative architectural pattern (HIIC)**

In order to make the operation of the MAPE-K loop proposed by ACon flexible enough for being adopted by modern SASs, a varying decentralization degree of the MAPE-K elements should be supported. Moreover, the communication and sharing of MAPE-K elements between different SASs should be supported. In order to formally describe how software components should behave in SACRE, we extended the notation described by Weyns et al. (2013) for decentralized control in SASs and propose the HIIC pattern (see Fig. 3). HIIC combines and extends the hierarchical and collaborative patterns described by Weyns et al. (2013).

The HIIC pattern consist of: a bottom layer, corresponding to the domain specific managed elements' logic; a middle layer, in charge of the adaptation of the domain specific managed elements placed on the bottom layer (e.g., evaluating their contextual requirements); and a top layer, for managing the adaptation of the middle layer operation in order to better support the managed elements (e.g., adapting the contextual requirements utilized by the middle layer, when runtime uncertainty is experienced). Moreover, we propose MAPE-K elements that communicate, if needed, with other MAPE-K elements of the same nature (e.g., a monitor with other monitors) and elements that can communicate with one or more MAPE-K elements of other types (e.g., a monitor with two analyze elements). In order to do that, we introduced cardinalities to the inter-components and intra-components interactions. This allows not only the communication and sharing of loop elements (addressing in this way **Chl2.2**), but the resilience of the loop in case an element of one type fails. We have also introduced cardinalities to the managing-managed subsystem interaction for indicating that one or more managed elements can be supported.

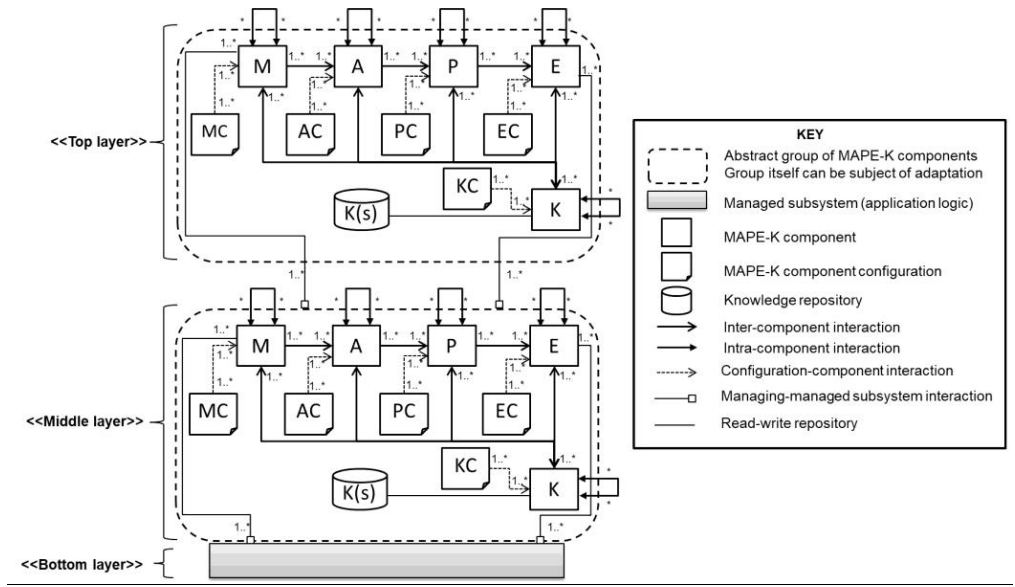

*Fig. 3. Hierarchical inter-intra collaborative pattern (HIIC)*

In order to address **Chl2.3**, i.e., freely vary the decentralization degree of the loop elements, we extended the notation with a *MAPE-K component configuration* and *Configuration-component interaction* terms. These configuration entities (MC, AC, PC, EC and KC in Fig. 3), contain all the knowledge that a loop element needs to perform its functionalities. Apart from enabling the decentralization, the explicit representation of elements' knowledge as a separate entity allows the later re-configuration of it and in consequence of the operation of its corresponding MAPE-K element without affecting the operation of the rest of the elements of the loop. Finally, we included the Knowledge base (K) as a loop component that manages the interactions with the Knowledge repository introduced by Weyns et al. (2013), and that is also part of the decentralization mechanism. An instantiation of this pattern will be provided later in this section as part of the reference architecture proposed by our approach SACRE.

### 5.3. SACRE design

SACRE is our proposal for addressing the four prevalent challenges affecting modern SASs that we have previously identified in this work in Section 2 (**Chl2.2**, **Chl2.3**, **Chl3.1** and **Chl3.2**). In this section, we place SACRE in a reference architecture (shown in Fig. 4) that maps the functionalities of the MAPE-K reference model proposed by ACon as well as new functionalities required for supporting the relations proposed by the architectural pattern HIIC, for decentralized and collaborative loops onto software components.

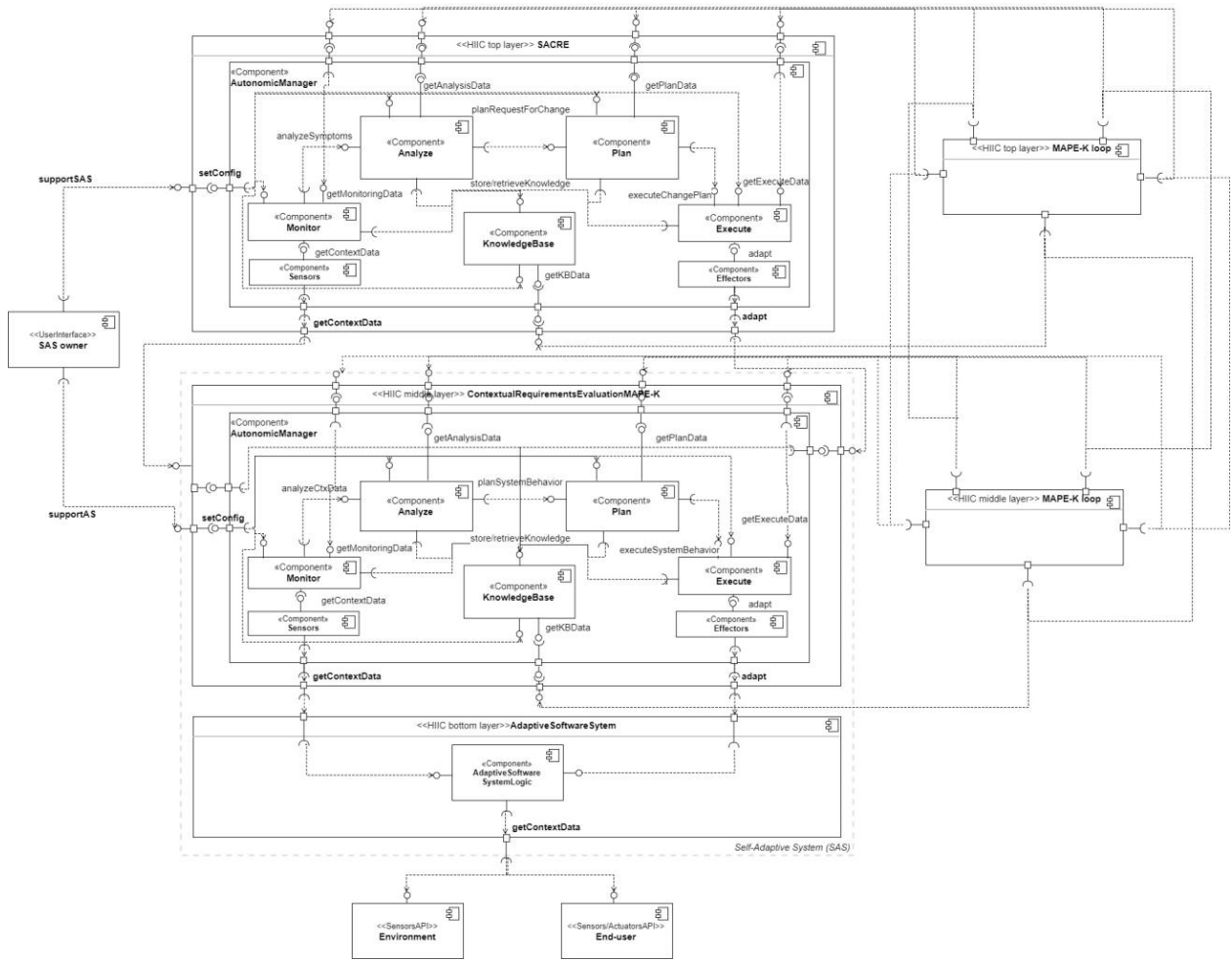

*Fig. 4. A reference architecture for modern SASs contextual requirements' adaptation with SACRE*

The architecture in Fig. 4 contains the following components:

- **SACRE**: HIIC top layer MAPE-K loop in charge of the operation of SACRE for adapting contextual requirements at runtime.
- **ContextualRequirementsEvaluationMAPE-K**: HIIC middle layer MAPE-K loop in charge of the adaptive systems' contextual requirements' evaluation.
- **AdaptiveSoftwareSystem**: HIIC bottom layer managed element that is adapted by the ContextualRequirementsEvaluationMAPE-K component.
- **SAS owner**: User interface for setting the configuration to the top and middle layer loops through the *supportSAS* and *supportAS*. This in turn is sent to each of the elements of the loops through their *setConfig* interfaces.
- **Environment and End-user**: These components provide context data to the AdaptiveSoftwareSystem through their *getContextData* interface. These data is later passed to the middle and top layer loops for further analysis.
- **Other MAPE-K loops**: middle and top layer black box MAPE-K loops (placed on the right side of Fig. 4) that exemplify how loop elements interact with other elements, of the same type, of other MAPE-K loops. The interaction is effectuated through the elements' provided (and required) interfaces *getMonitoringData*, *getAnalysisData*, *getPlanData*, *getExecuteData* and *getKBData*. The interaction between elements of different types, for sharing their functionalities with other loops supporting SASs, can be achieved by exposing the interfaces, used for the normal operation of the loops (i.e., *analyzeSymptoms*, *planRequestForChange*, *executeChangePlan* and *store/retrieveKnowledge* in SACRE loop and *analyzeCtxData*, *planSystemBehavior*, *executeSystemBehavior* and *store/retrieveKnowledge* in ContextualRequirementsEvaluationMAPE-K loop) to external components (as it is exemplified with the interfaces *getMonitorinData*, *getAnalysisaData*, etc.).

As it can be noticed, for the sake of simplicity, in Fig. 4 we provide MAPE-K loops with only one element of each type (i.e., one Monitor, one Analyze, etc.) and not shared elements. However, thanks to the adoption of the HIIC architectural pattern and the components' interfaces we propose, different combinations of top and middle layers loops are possible. In the rest of this section, we describe the operation of SACRE. Then, in Section 6 we will discuss the ContextualRequirementsEvaluationMAPE-K and AdaptiveSoftwareSystem components.

SACRE adopts the communication and knowledge decentralization ideas of HIIC. The configuration knowledge, provided by the SASs owners through the *supportSAS* interface in Fig. 4 is given in the form of a set of policies. First, owners provide policies in a preparation phase in order to enable SACRE components to support their SASs, and later when desired, at runtime. One MAPE-K element can manage more than one policy (e.g., for supporting more than one SAS or set of contextual requirements) and a policy can be used for more than one loop element of the same type (e.g., two monitors can share a policy for redundancy purposes). Concretely, the policies considered in SACRE are:

- **Autonomic manager policy**. This policy contains the MAPE-K loop structural details required by the SAS(s) to be supported and the rest of MAPE-K elements' policies. It is introduced as a unique input to SACRE in order to configure a complete MAPE-K loop (centralized, decentralized, with one or more elements of one type, etc.) for serving a specific (set of) SAS(s).
- **Monitor policy**. This policy is delivered to the Monitor type element and contains the following configuration variables: environmental variables to monitor, pre-processing parameters (e.g., for normalizing measuring values) and thresholds for detecting anomalous variables values (e.g., because a sensor decalibration).
- **Analyze policy**. This is the policy for the Analyze type element and contains the following configuration variables: data mining algorithm and tool to be used, algorithm environmental input variables and expected output statistical measures.
- **Plan policy**. This policy is delivered to the element of type Plan and describes the configuration variables that contain the acceptance thresholds for the data mining algorithm resulting measures.
- **Execute policy**. This policy is for the Execute type element and contains the configuration variable that stores the supported SAS(s) id(s) in order to dispatch correctly requirements' adaptations.
- **Knowledge base policy**. Finally, the policy for the Knowledge base type element contains the variables: frequency for indicating SACRE's inter-iterations time, minimum number of iterations experiencing uncertainty before triggering a Symptom and parameters for persistence (e.g., which environmental variables have to be persisted, in which format, etc.).

As the goal is to use the implementations of SACRE in different domains, we expect new configuration variables to be required and policies to be extended. Now, we introduce the communication mechanisms employed by SACRE for decoupling the elements of the loop at the same time as guarantying its correct operation.

The sequence diagram presented in Fig. 5, provides the details of the initialization process of SACRE (top layer loop in Fig. 4). In order to correctly operate, independently of its decentralization level, SACRE requires the configuration information of the SASs to support, as well as a complete MAPE-K loop with at least one instance of each type of element (M, A, P, E and K). In order to ensure that these two conditions are always met at runtime, we have design a coordination role that we assigned to the Automatic manager (component containing the MAPE-K loop elements in Fig. 4) which its only task is the fulfillment of these two conditions. In normal operation, this component does not intervene and the MAPE-K elements communicate to each other independently. However, if structural or configuration changes are required in the MAPE-K loop, the Autonomic manager is the one responsible of coordinating these changes (using the Automatic manager policy introduced before in this section).

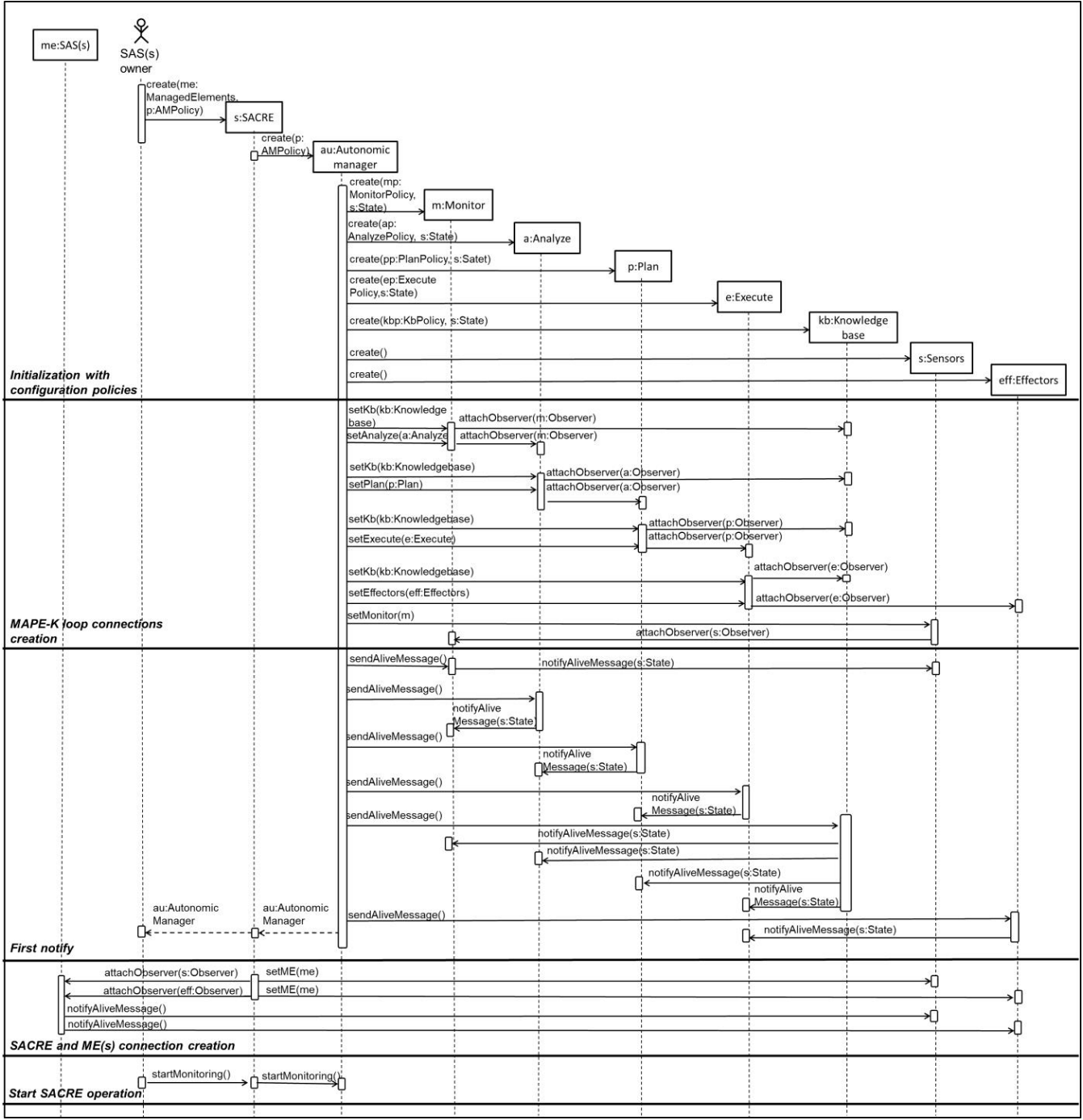

*Fig. 5. SACRE setup process*

The setup process of SACRE can be divided in five main phases:

1. **Initialization with configuration policies**. The process starts with the initialization of SACRE by the SAS(s) owner. The owner provides the Autonomic manager policy and the SAS(s) to support as parameters. Then the Automatic manager is created which in turn creates the MAPE-K elements (including Sensors and Effectors) as observable components (following the observer-observable pattern (Gamma et al., 1995)) and assigns them their corresponding policy. In the diagram of Fig. 5, we consider one instance of each MAPE-K element type for simplicity but different configurations, as explain before in this section, are possible. The MAPE-K elements have an attribute *State* used for indicating elements' health (e.g., OK if everything is working correctly, NO-OK, otherwise) which is initialized as OK.

2. **MAPE-K loop connections creation**. In this phase, the Autonomic manager creates the connections among the loop elements and the elements in turn attach themselves as observers to the corresponding elements. For instance, when the Autonomic manager sets an Analyze element to a Monitor element, for establishing their connection, the Monitor in turn attaches itself as observer of the Analyze element. In this way any State change of observable components is immediately detected by the components that have to interact with them. This can be later used for notifying loop anomalies, faults or re-configuration requirements.
3. **First notify**. In this step, the Autonomic manager requests all elements to communicate with their observables in order to verify the health of the complete loop. This step finalizes the create call.
4. **SACRE and ME(s) connections creation**. This phase consists in the connection of SACRE (the Autonomic manager with the MAPE-K elements just created) with the ME(s) to support, i.e., the SAS(s). The ME(s) is (are) set to the Sensors and Effectors interfaces. Then, the Sensors and Effectors are attached as observers of the ME(s). A first notify from the ME(s) finalizes this phase.
5. **Start SACRE operation**. In this last step, the SAS(s) owner starts SACRE and it in turn initiates the monitoring task.

For finalizing the description of our approach, in Fig. 6, an overview of the operation of SACRE in the form of a sequence diagram is provided. There the interaction between the MAPE-K elements, their policies and the ME(s) is detailed. The diagram exemplifies the following contextual requirements' adaptation process:

1. In the first loop, the SAS(s) (or ME(s)) continuously send context data to the Sensors interface. Thus, Sensors maintain up-to-date knowledge about the environmental variables.
2. In the second loop, the Monitor element senses context data from Sensors at the policy given frequency, stores the data in the Knowledge base and determines requirements affected by uncertainty using runtime and policy information. If uncertainty has been experienced for a certain number of iterations (indicated in policies), the Monitor sends a Symptom to the Analyze element. The Analyze element gets the historical context data from the Knowledge base and analyzes it using data mining. If significant results are obtained for a certain number of iterations (indicated in policies), the Analyze sends a Request for Change to the Plan element. Then, the Plan element determines if an adaptation should be executed, according to the measures acceptance thresholds indicated in its policy. If it is the case, the Plan sends a Change Plan to the Execute element. Finally, the Execute element enacts the new requirements' operationalization in the SAS(s) through Effectors.

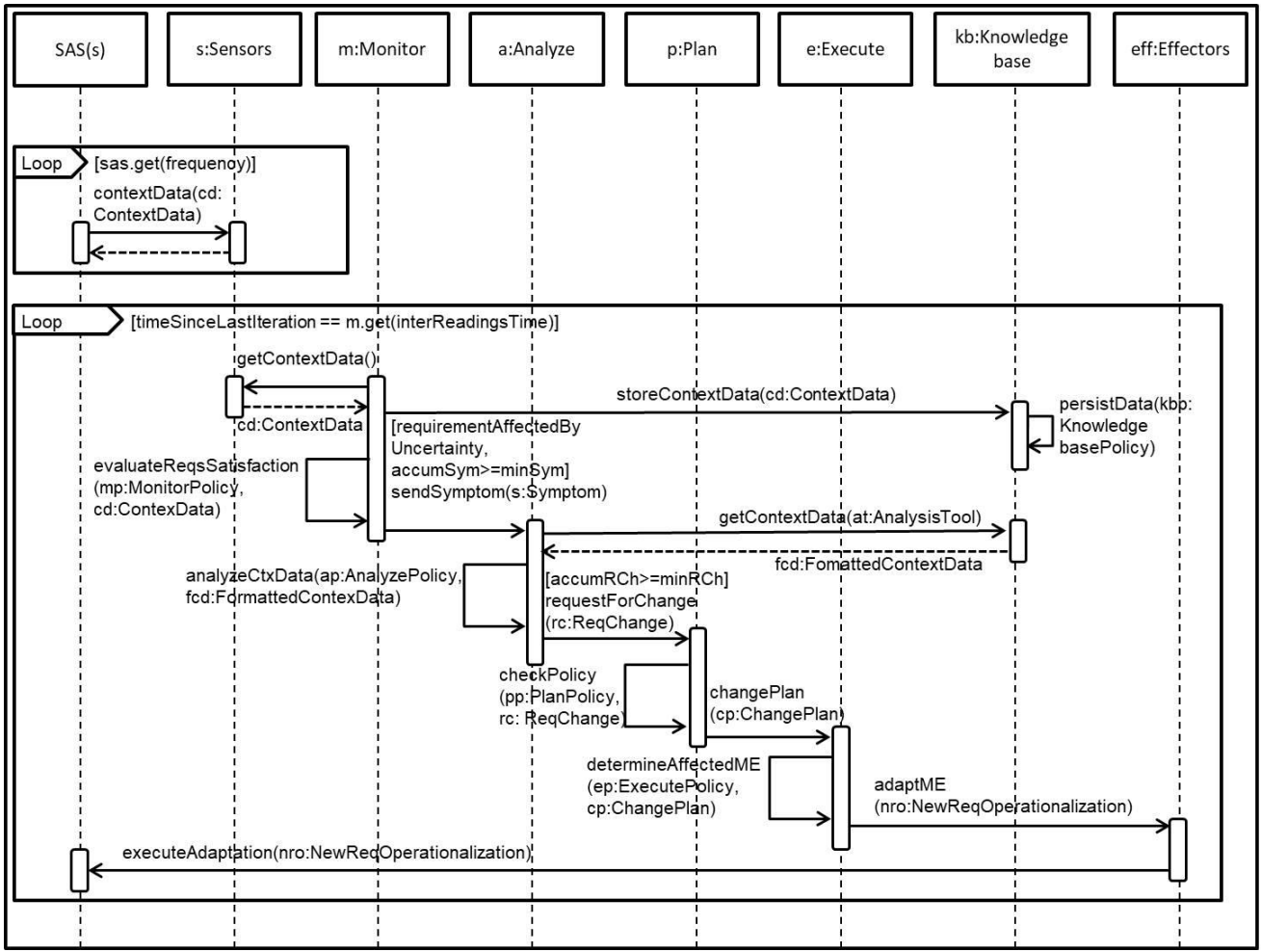

*Fig. 6. Contextual requirements adaptation in SACRE*

Next sections are dedicated to evaluate SACRE. As mentioned before in Section 1, the evaluation is performed in the extremely demanding domain of smart vehicles, thus in next section we provide some details about this domain.

## 6. Application of SACRE in the domain of smart vehicles

Smart (or intelligent) vehicles are systems capable of sensing contextual data (e.g., from the driver, the environment, the vehicle itself) and making decisions based on these data (e.g., turn on an alarm, activate self-driving functionality). These systems have become increasingly popular in the automotive industry. In consequence, the interest of the research community in this domain has increased steadily. Smart vehicles have brought several societal benefits, for instance, improving drivers' safety, optimizing fuel consumption, improving driver's experience and comfort. At the same time, their control systems need to face challenging characteristics such as self-adaptability, self-configurability, high-level of human interaction, uncertainty, what make this domain still subject of research.

In this work, we use contextual requirements for describing a particular functionality of a smart vehicle involving self-adaptation capabilities. It consists of the detection and support of drowsy drivers. We use this functionality as an example for evaluating SACRE.

**6.1. Contextual requirements for detecting and supporting drowsy drivers**

Many of the road accidents are occurring due to driver fatigue (i.e., driver drowsiness or driver sleepiness). Sleepiness reduces the concentration, activeness, alertness, and vigilance of the driver and it makes the driver to take slow decisions and sometimes no decisions at all. Drowsiness affects the mental alertness and decreases the ability of the driver to operate a vehicle safely, increasing the risk of human error that could lead to fatalities and injuries. Hence, to increase the road safety, there is a need to address this issue to avoid accidents by alerting the driver (Krishnasree et al., 2014). In order to do so, the state of drowsiness and alertness of the driver should be effectively monitored (Sahayadhas et al., 2012).

Drivers can experience different levels of drowsiness and alertness, from drowsy to dangerously drowsy and finally sleeping. In order to prevent accidents, different mechanisms for alerting and supporting drivers have been used, for example, in previous work (Lee et al., 2004) auditory and seat-based vibration warnings have been proposed to mitigate driver distraction. Other examples are lane keeping assistance systems and lane departure avoidance systems studied by many researchers (Lee et al., 2014). In our running example, in order to detect and support drowsy drivers at different stages, we have defined three levels of drowsiness with three different actuators to support each of these levels. In order to be supported by SACRE, we have modeled them as contextual requirements (see *Table 3*).

*Table 3. Contextual requirements*

| ID | Context | Behavior |
|---|---|---|
| $cr_1$ | Driver is drowsy | Activate seat-vibration alarm |
| $cr_2$ | Driver is dangerously drowsy | Activate sound-light alarm |
| $cr_3$ | Driver is sleeping | Activate lane keeping support |

The different drowsiness levels correspond to the <<*context*>> of the requirement, while the activation of the actuators corresponds to the <<*expected system behavior*>>. The satisfaction of the contextual requirements consists of the execution of the corresponding expected system behavior when a context holds at runtime. In order to continuously evaluate contexts, we use sensor data. Thus, situations such decalibration of sensors or faults can cause contextual requirements dissatisfaction. In the next section, we describe the set of sensors used in our running example for the operationalization of the contextual requirements' context from *Table* 3.

**6.2. Sensors**

According to previous work (Sahayadhas et al., 2012), there are three types of measures that have been used widely for monitoring drowsiness:

- **Vehicle-based measures**. These include deviations from the lane position, movement of the steering wheel, pressure on the acceleration pedal, etc., once crosses a specified threshold, it indicates a significantly increased probability that the driver is drowsy.
- **Behavioral measures**. For example, yawning, eye closure, eye blinking, head position, monitored through a camera.
- **Physiological measures**. Namely electrocardiogram (ECG), electromyogram (EMG), electrooculogram (EoG) and electroencephalogram (EEG).

For the development of an efficient drowsiness detection system, the strengths of the various measures should be combined into a hybrid system (Sahayadhas et al., 2012). In our running example, we use three different sensors for obtaining measures from the three types mentioned above. Below, we describe each of the sensors used in our example to represent the three kinds of measures:

- **Steering wheel pressure sensor**: The lack of hands or only one hand on a steering wheel could be an indication of a drowsy driver (Lisseman et al., 2015). Based on the patent presented by Lisseman et al., we consider a steering wheel pressure sensor (green triangles in Fig. 7) for continuously obtaining the number of driver's *hands on the steering wheel* (*hosw*).

- **Driver's vigilance level camera sensor**: Eyes closure during long periods and frequently non-frontal face position are clear symptoms of driver fatigue (Bergasa et al., 2006). Camera-based sensors for monitoring drivers in real time, as the prototype presented in previous work (Bergasa et al., 2006), have been proposed for extracting behavioral measures such the ones mentioned before. In this work, we consider a camera sensor (yellow circle in Fig. 7) able to report: the *eyes state*, indicated by the percentage of the driver's pupils that is visible; and the *face position* measure, with two possible values frontal and non-frontal. Then, as in previous work (Bergasa et al., 2006), using the eyes state variable we calculate the *per*cent eye *clos*ure (*perclos)* measure, consisting in the percentage time where driver's pupils have been less than 20% visible.

- **Electrocardiogram sensor**: As driving fatigue develops heart rate slows, triggering a series of events (i.e., blood pressures go down, poor circulation and finally hypoxia in brain) that induces drowsiness and loss of concentration (Liang et al., 2009). More and more non-intrusive electrocardiogram sensors, as the one presented by Wartzek et al. (2011), are developed for being incorporated in vehicles in order to continuously monitor drivers' heart rate. Based on Wartzek et al. prototype, in this work we consider an electrocardiogram sensor (blue squares in Fig. 7) for obtaining the physiological measure *hearth beats per minute (hbpm)*.

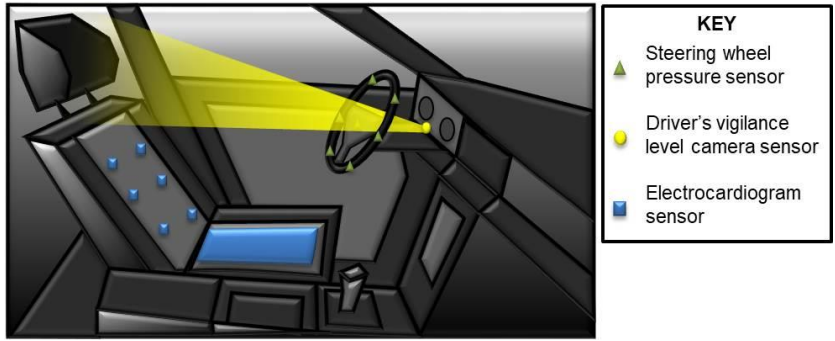

*Fig. 7. Smart vehicle sensors (i.e., steering wheel pressure sensor,
driver's vigilance level camera sensor and electrocardiogram sensor)*

As explained in Section 5.1, the measures obtain by the sensors (i.e., environmental variables) listed above (i.e., *hosw, perclos, face position* and *hbpm*) are combined by expression operators (i.e., relational, arithmetic, and logical) in order to operationalize the contextual requirements' contexts. For example, a context operationalization of $cr_1$ (i.e., driver is drowsy, see *Table* 3), could be:

$$perclos >= 15 \text{ AND } hbpm >= 67 \text{ AND } hbpm <= 72$$

In the next section, we describe the actuators used in our running example for executing the contextual requirements' expected behaviors as defined in *Table* 3.

### 6.3. Actuators

When drowsiness is detected, smart vehicles may use feedback to warn the driver. Such feedback can be, for example, a warning sound, voice, light or vibration (Lisseman et al., 2015). When driver's drowsiness level reaches high values, more sophisticated actuators such as lane keeping systems may support them better. Below, we list the set of actuators we consider in this work:

- **Seat-vibration alarm**: Graded seat-vibration alarm has been perceived as a trusted (and less annoying) actuator for warning drivers (Lee et al., 2004). In this work, we consider a *seat-vibration alarm (orange circles in* Fig. 8*)* based on the prototype presented by Lee et al., for supporting drowsy driver context (see $cr_1$ in *Table* 3).

- **Sound-light alarm**: Substantial research shows that complementing visual cues with redundant cues in another sensory mode speeds people's reaction time. A common combination of sensors is the use of visual and auditory displays (Lee et al., 2004). Thus, in our running example, we use a *sound-light alarm* actuator (red squares in Fig. 8) for alerting dangerously drowsy drivers (see $cr_2$ in *Table* 3).

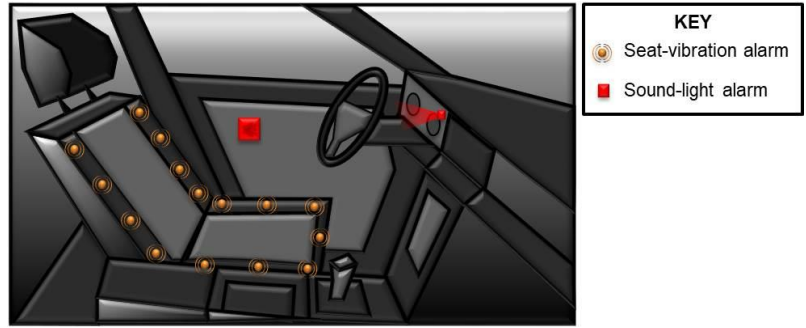

*Fig. 8. Smart vehicle actuators (seat-vibration and sound-light alarms)*

- **Lane keeping support**: In order to prevent accidents caused by fatigue drivers, many researchers have focused on studying lane keeping assistance systems and lane departure avoidance systems (Lee et al., 2014). These systems are characterized to be activated punctually when critic situations occur (e.g., when a driver falls asleep). We consider a *lane-keeping support* system for responding to the *driver is sleeping* context (see $cr_3$ in *Table* 3). A technology similar to the one shown by BMW (2016) and Toyota Motor Sales (2016) could be considered. Fig. 9 provides an illustration of this actuator.

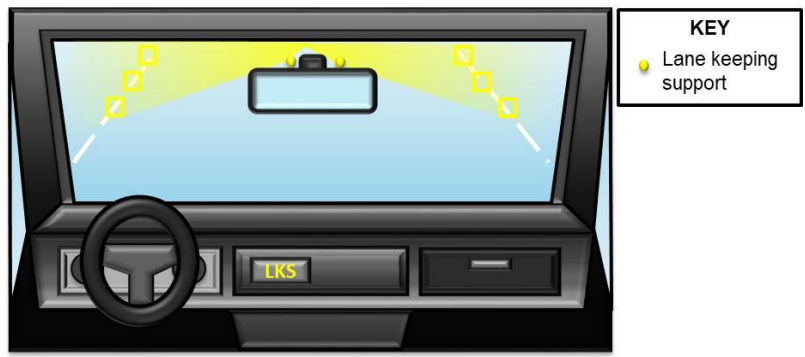

*Fig. 9. Smart vehicle actuators (lane keeping support system)*

Actuators can be turned off by the driver after they have been activated by the system (for satisfying the contextual requirements) or disabled when the driver wants to keep them off. Particularly, the lane keeping support actuator can also be turned on by the driver. The actions of turning on/off and disabling actuators, triggered by the driver, result in candidate adaptations of the system's contextual requirements. Therefore, it will happen that an adapted requirement triggers the system to activate an actuator at some point of time, while in the past it was not the case, or the other way around.

## 7. Implementation of SACRE for supporting smart vehicles

In the smart vehicles domain, system's contextual requirements are constantly affected by uncertainty due to the inability to anticipate every single possible environment event (e.g., weather conditions, road in bad state, driver decisions) and to the need to configure the behavior of the system to the driver's preferences and abilities, which can also vary due to the context and evolve over time. This issue challenges automotive industries for providing a generic software solution (to be adopted for any of their smart vehicles) able to adapt to each of their clients (i.e., drivers) at runtime, without the need of any further maintenance (e.g., sensors recalibration). These needs make this domain an excellent evaluation example for our improved approach, SACRE.

The smart vehicle operation of our running example follows the reference architecture proposed in Section 5. The MAPE-K loop of the smart vehicle corresponds to the ContextualRequirementsEvaluationMAPE-K component shown in Fig. 4, while the adaptive smart vehicle logic, to the AdaptiveSoftwareSystem component. The loop receives the environment variables monitored by the smart vehicle from external systems (Environment and End-user components in Fig. 4), determines which requirements' <<context>> holds, and activates the corresponding <<expected system behavior>>. Moreover, it sends sensor data to the Monitor of SACRE and receives requirements' re-operationalizations from SACRE's Execute component. Contextual requirements are stored in the Knowledge base of the ContextualRequirementsEvaluationMAPE-K loop. The configuration policies of the smart vehicle loop contain application-specific configuration variables that may vary from one application to another. In this specific implementation, we do not consider the interaction of the top and middle layer loops with other MAPE-K loops. Moreover, the loops are implemented with only one element of each type.

The smart vehicles domain belongs to the embedded devices environment in which resources (e.g., CPU, memory) are very limited and the demand is really high. Among other factors, the need to reduce the time-to-market and the development complexity of the applications for embedded devices has encouraged the creation of friendly programming languages such as Java Micro Edition (Java ME), which we use in our running example. In Fig. 10, we provide details about how Java ME and other well-known technologies have been used for simulating a smart vehicle and implemented SACRE's adaptation feedback loop. Moreover, the diagram provides details about the interaction between the software modules.

The implementation consists of two main parts:

1) **SACRE.**
    - **MAPE-K loop**. In our example, we are simulating an adaptation loop installed in a smart vehicle (i.e., an embedded device). For this reason, we have used Java ME 8.1 for implementing both the MAPE-K elements (i.e., monitor, analyze, plan, execute, knowledge base, sensors and effectors) that compose the core of SACRE and their corresponding policies' managers (see Fig. 10). In this implementation, the policies have been created at design-time as *.properties* files and stored in the resources folder of the project. As it is shown in Fig. 10, we have used asynchronous communication between the Monitor and the Analyze and between the Monitor and the Knowledge base modules. By experience, we have noticed that these are the more work-intensive modules and with synchronous communication they affect the response time of the whole loop. Thus, with the asynchronous communication implemented they operate independently, at their own pace (i.e., monitoring

and calculate requirements' satisfaction, analyze data and persist context data). The Knowledge base module persist the context data in the form of *.arff* files in the File System for being later used by the data mining tool. More details about these files will be provided in the Analysis tool.

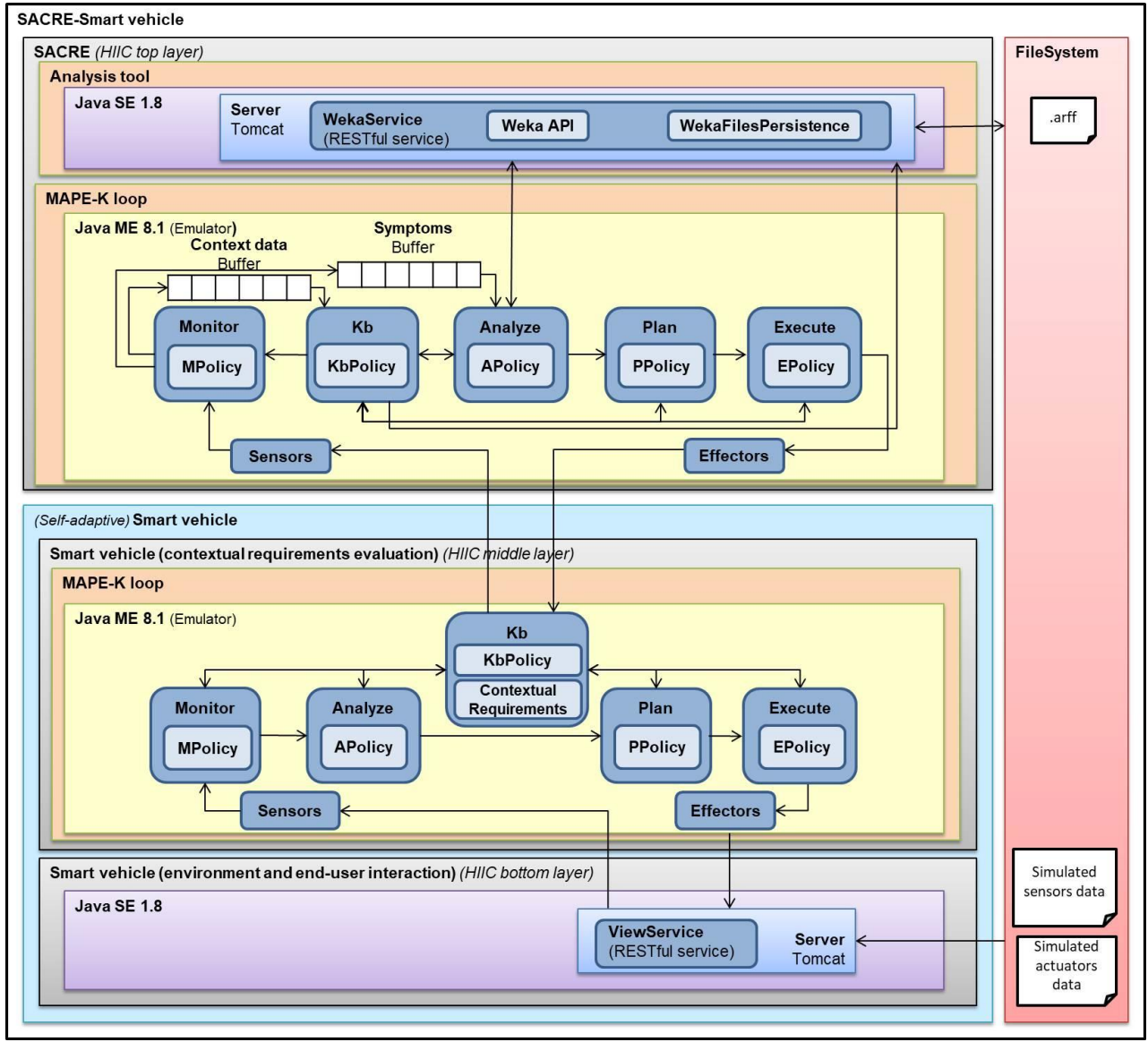

*Fig. 10. Implementation of SACRE for the smart vehicles domain*

- **Analysis tool**. As mentioned before in this section, the smart vehicles belong to a domain in which resources are very limited. Thus, an appropriate data mining algorithm should be considered for performing the analysis. In our previous work (Knauss et al., 2016), the rule-based JRip algorithm (Kotsiantis, 2007; Tan et al., 2005) and the Waikato Machine Learning tool, Weka (Machine Learning Group at the University of Waikato, 2016) have been used for supporting an application with similar restrictions. For this reason, in this running example we use the same algorithm. The Analysis tool component shown in Fig. 10 is in charge of using the Weka tool Java API for applying the JRip algorithm to the SACRE's context data at runtime.

It also offers a second method for supporting the Knowledge base with the persistence (i.e., creation, deletion and update) of the context data, in the form of *.arff* files, in the File System. Later, these files are retrieved by the method using the Weka API. The *.arff* format is the format used by Weka and it consists in a file with a header and a body. The header contains the list of the context variables and their type, as well as the variable to analyze. The actual context data is then provided in the body. Since Java ME was not compatible, we used Java SE (Java Platform, Standard Edition) 1.8 for implementing this component. In order to enable the communication between the Analysis tool and the MAPE-K components, the tool has been implemented as a RESTful service. The separation of the data mining algorithm implementation and the MAPE-K Analysis module allows future implementations to reuse the Analysis tool service or substitute it for investigating other algorithms such the ones discussed in previous works for similar purposes (Knauss et al., 2016; Rook, 2014; Rook et al., 2014).

2) **Smart vehicle**.
   - **MAPE-K loop**. The MAPE-K loop in charge of the contextual requirements evaluation has been also implemented in Java ME 8.1. As it is shown in Fig. 10, contextual requirements are stored in the Knowledge base which is in charge of updating them when an adaptation request from SACRE is received. The initial set of requirements is provided also as a *.properties* configuration file. Then, it is stored in memory and when an adaptation request is received from SACRE, the memory variables are updated (i.e., the original file is not overwritten). The rest of the loop elements are notified when the contextual requirements have been adapted, so they can evaluate the new set.
   - **Interaction module**. We included a module in the smart vehicle for simulating the interaction of the driver with the smart vehicle. This module reports sensors' and actuators' data. The module has been implemented in Java SE 1.8. The communication with the smart vehicle MAPE-K (implemented in Java ME) is done through a RESTful service (View service in Fig. 10). Sensors' and actuators' data is simulated and read by the smart vehicle view module from the File System. These files have been created at design-time.

The source code of this implementation as well as more details about its construction, artifacts and instructions of usage are available at https://github.com/edithzavala/sacre-sv. In next section we describe the evaluation of SACRE using this implementation.

## 8. Evaluation of SACRE

This section describes the evaluation of SACRE in terms of its response time and accuracy when adapting contextual requirements' operationalization in the presence of runtime uncertainty. The evaluation is conducted using our running example in the domain of smart vehicles. Particularly, we focused on the functionality of detecting and supporting drowsy drivers modeled by the contextual requirements presented in *Table* 3. For the evaluation we have used the implementation of SACRE presented in Section 7. In order to asses SACRE, we have exposed the approach to different uncertainty scenarios (e.g., a sensor fault). The evaluation has been performed in real-time using a simulated environment. Concretely, we have used an Intel Core$^{TM}$ 2 Duo CPU P7350 @ 2.00 GHz, with 3,0GB RAM for running the evaluation. In the remainder of this section, we describe the evaluation process and the threats to validity we identified for this evaluation. In summary, the evaluation process consists in the following activities:

- *Preparation for the application of SACRE*. This activity consists in determining the configuration variables' values for the smart vehicle (logic) and the SACRE's loop elements policies as well as the uncertainty scenarios in which SACRE will be evaluated.
- *Adaptation of contextual requirements affected by uncertainty*. This activity consist in the execution of SACRE in the different uncertainty scenarios for triggering the adaptation of the contextual requirements' affected by uncertainty.
- *Analysis of the results*. In this activity, we analyze the results of the requirements' context adaptation 1) exploring SACRE's resulting response time (since a Symptom is triggered until the adaptation is enacted) and 2) statistically analyzing the data mining algorithm.

## 8.1. Preparation for the application of SACRE

In *Table* 4 and *Table* 5, we provide the configuration variables' values we have set in this evaluation for SACRE and the smart vehicle respectively. Instructions about where these configuration variables can be tuned in our system and which values they can take are provided in https://github.com/edithzavala/sacre-sv (*getting started* section).

*Table 4. SACRE's policies*

| Policy | Configuration variable | Value |
|---|---|---|
| **Autonomic manager** | MAPE-K Structure | Default. *The default value refers to a MAPE-K loop with one instance of each type of loop element.* |
| | MAPE-K polices | Monitor, Analyze, Plan, Execute and Knowledge base policies |
| **Monitor** | Monitoring variables | perclos, facePosition, heartBeatsPerMinute, handsOnSteeringWheel |
| | Monitoring variables normalization max | perclos Max = 100, facePositionMax = 1 heartBeatsPerMinuteMax = 120, handsOnSteeringWheelMax = 2 |
| | Monitoring variables normalization min | perclos Min = 0, facePositionMin = 0, heartBeatsPerMinuteMin = 0, handsOnSteeringWheelMin = 0 |
| | Pre-processing functions | perclos = calculate(eyesSate), facePosition = -, heartBeatsPerMinute = -, handsOnSteeringWheel = - |
| | Monitoring variables min | perclos Min = 0, facePositionMin = 0, heartBeatsPerMinuteMin = 0,3, handsOnSteeringWheelMin = 0 |
| | Monitoring variables max | perclos Max = 1, facePositionMax = 1 heartBeatsPerMinuteMax = 1, handsOnSteeringWheelMax = 1 |
| **Analyze** | Analysis variables | perclos, facePosition, heartBeatsPerMinute, handsOnSteeringWheel |
| | Data mining algorithm | JRip |
| | Data mining tool | Weka |
| | Data mining expected output | Rules, Precision, Recall, fMeasure. |
| | Min analysis iterations | N/A. *This parameter has been indicated in source code. Since no experimental evidence is available for setting this parameter we configure it as 0 iterations (except for uncertainty case 2(a, b and c) in Table 2 for which we set 3 iterations because in that case data mining is not used and we want to avoid requirements' adaptation triggered by isolated uncertainty situations).* |
| **Plan** | Data mining measures | Precision, Recall, fMeasure |
| | Data mining measures min | PrecisionMin = 0,95, RecallMin = 0,95, fMeasureMin = 0,95 |
| **Execute** | Managed element(s) | N/A. Indicated in code (smart vehicle). |
| **Knowledge base** | SACRE frequency | 14,28 iterations per second. *By experience, this is the highest frequency rate SACRE can reach in the machine used for the evaluation. Authors from previous work (Jiménez-Pinto & Torres-Torriti, 2013) in the domain of smart vehicles, determined that a minimum rate of 5-10 iterations per second is required for correctly detecting drivers fatigue. Thus, this frequency value is good and supported by the literature.* |
| | Min uncertainty iterations | 3 iterations. *There is no evidence for setting this parameter, thus we inverted the analysis policy configuring 3 iterations for the uncertain cases requiring data mining and 0 (indicated in code) for uncertainty situations of case 2 (a, b and c in Table 2).* |
| | Variables to persist | perclos, facePosition, heartBeatsPerMinute, handsOnSteeringWheel, cr1ExpectedBehaviorState, |

| | | cr2ExpectedBehaviorState, cr3ExpectedBehaviorState |
|---|---|---|

*Table 5. Smart vehicle's policies*

| *Policy* | *Configuration variable* | *Value* |
|---|---|---|
| **Autonomic manager** | MAPE-K Structure | Default. The default value refers to a MAPE-K loop with one instance of each type of loop element. |
| | MAPE-K polices | Monitor, Analyze, Plan, Execute and Knowledge base policies |
| **Monitor** | Monitoring variables | eyesState, facePosition, heartBeatsPerMinute, handsOnSteeringWheel |
| | Monitoring variables normalization max | eyesStateMax = 1, facePositionMax = 1 heartBeatsPerMinuteMax = 120, handsOnSteeringWheelMax = 2 |
| | Monitoring variables normalization min | eyesState Min = 0, facePositionMin = 0 heartBeatsPerMinuteMin = 0, handsOnSteeringWheelMin = 0 |
| | Pre-processing functions | N/A. Indicated in code (perclos = calculate(eyesState)). |
| **Analyze** | Contextual requirements' context | ctx1: Driver is drowsy<br>ctx2: Driver is dangerously drowsy<br>ctx3:Driver is sleeping |
| | Contextual requirements' context operationalization variables | var1: perclos<br>var2: facePosition<br>var3: heartBeatsPerMinute (hbpm)<br>var4: handsOnSteeringWheel (hosw) |
| | Contextual requirements' context operationalization | ctx1Oper = *perclos>=0,15 AND hbpm<=0,60 AND hbpm >=0,56*<br>ctx2Oper = *perclos>=0,21 AND facePosition=1 AND hbpm<=0,55 AND hbpm>=0,46*<br>ctx3Oper = *perclos >0,30 AND facePosition=1 AND hbmp<=0,45 AND hosw <1* |
| **Plan** | Contextual requirements' expected system behavior | beh1: Activate Seat Vibration,<br>beh2: Activate Sound/Light Alert<br>beh3: Activate lane keeping support |
| | Contextual requirements | cr1: ctx1, beh1<br>cr2: ctx2, beh2<br>cr3: ctx3, beh3 |

| **Execute** | Managed element(s) | N/A. Indicated in code (adaptive vehicle). |
|---|---|---|
| | MAPE-K loop frequency | 20 iterations per second. |
| **Knowledge base** | Contextual requirements' operationalization update functions | ctx1Oper = process(ctx1*SACRENewOper*) <br> ctx2Oper = process (ctx2*SACRENewOper*) <br> ctx3Oper = process (ctx3*SACRENewOper*) |

In order to evaluate SACRE we have design six uncertainty scenarios ($us_1$, $us_2$, $us_3$, $us_{4a}$, $us_{4b}$ and $us_5$ in *Table 6*). Each scenario focuses on a specific uncertainty case (from *Table* 2) that at certain point (indicated by the number of iterations in *Table 6*) affects one or more contextual requirements (from *Table* 3), triggering the adaptation of them.

*Table 6. Uncertainty evaluation scenarios*

| *Uncertainty scenario* | *Uncertainty case* | *SACRE iterations* |
|---|---|---|
| $us_1$ | cr1 affected by uncertainty case 3 <br> *(vibration alarm disabled)* | 1.000 |
| $us_2$ | cr2 affected by uncertainty case 3 <br> *(sound-light alarm disabled)* | 15.000 |
| $us_3$ | cr3 affected by uncertainty case 3 <br> *(lane keeping support disabled)* | 30.000 |
| $us_{4a}$ | cr2 and cr3 affected by uncertainty case 2b <br> *(driver's vigilance level camera sensor reports facePosition variable values out of thresholds)* | 45.000 |
| $us_{4b}$ | cr1, cr2 and cr3 affected by uncertainty case 2c <br> *(driver's vigilance level camera sensor reports facePosition variable values within thresholds again)* | 60.000 |
| $us_5$ | cr3 affected by uncertainty case 4 <br> *(lane keeping support manually activated)* | 75.000 |

In this implementation of SACRE, more iterations mean more context data persisted in the Knowledge base to analyze. Thus, we have trigged the uncertainty cases in the scenarios at different points in order to explore how the accuracy and response time of SACRE is impacted by the different amount of data available to analyze. We expect this parameter to do not impact in the scenarios in which data mining analysis is not required, this are the scenarios affected by uncertainty case 2 (b and c). In SACRE, as in ACon, the analysis of this uncertainty case consists in the automatic addition or removal of the sensor variable from the set of active variables, thus no pattern analysis in historical data is considered.

The uncertainty scenarios have been simulated by the smart vehicle view module introduced in Section 7. This module reads two comma-separated values (cvs) files containing simulated sensors' and actuators' data respectively. The content of the files varies from one uncertainty scenario to another, in https://github.com/edithzavala/sacre-sv we provide examples of these files which can be directly tried in our system or modified for evaluating other scenarios.

## 8.2. Adaptation of contextual requirements affected by uncertainty

Having determined the configuration of SACRE and the smart vehicle as well as the uncertainty scenarios in which our approach will be assessed, we proceed to the execution of the evaluation. In order ensure the reliability of our results and evaluate the robustness of SACRE for smart vehicles, we decided to replicate the execution of the uncertainty scenarios several times. For calculating a correct number of replications, we have used the formula (Berenson & Levine, 1996):

$$n = \frac{Z_\alpha^2 N p q}{e^2 (N-1) + Z_\alpha^2 p q}$$

Where:

- $n$: is the number of replications we require (i.e., sample size).
- $Z_\alpha$: is a constant that depends on the confidence level we desire for our evaluation. We selected a typically used 95% of confidence level which is a $Z_\alpha$ of 1,96.
- $N$: is the total population size. In our case this is the total number of executions we expect for the system, i.e., the smart vehicle with SACRE integrated. Considering a vehicle lifespan of 15 years (Baumhöfer et al., 2014) and a twice-daily use, we set a total population size of 10,950 executions.
- $e$: is the sample error we accept for this evaluation. We considered a 0,1.
- $p$ & $q$: are the probability of success and failure respectively. We used the typical value of 0,5 for each of them.

Given the formula and variables' values presented above, we obtained a *n* of 95,21 replications. Based on this result, we decided to run 100 replications for each uncertainty scenario.

We then executed the uncertainty scenarios. Fig. 11 shows the normalized values that the sensors' variables have taken in each of the uncertainty scenarios over execution time, represented each in a separate sub-graph. The *x*-axis of each sub-graph shows the number of SACRE iteration, the *y*-axis the normalized sensor variable value. As mentioned before in Section 8.2, these values have been simulated and introduced to the system through the smart vehicle view component. On the other hand, Fig. 12 shows the actuators' state (0 for inactive, 1 for active) during the execution of each scenario, each actuator is represented in a separate sub-graph. The *x*-axis of each sub-graph shows the number of SACRE iteration, the *y*-axis the actuator state. Actuators' state has also been simulated and manipulated for triggering adaptations (e.g., deactivate an actuator when it should be active). Moreover, Fig. 11 and Fig. 12 show exactly when the uncertainty cases in *Table 6* have been triggered in each scenario.

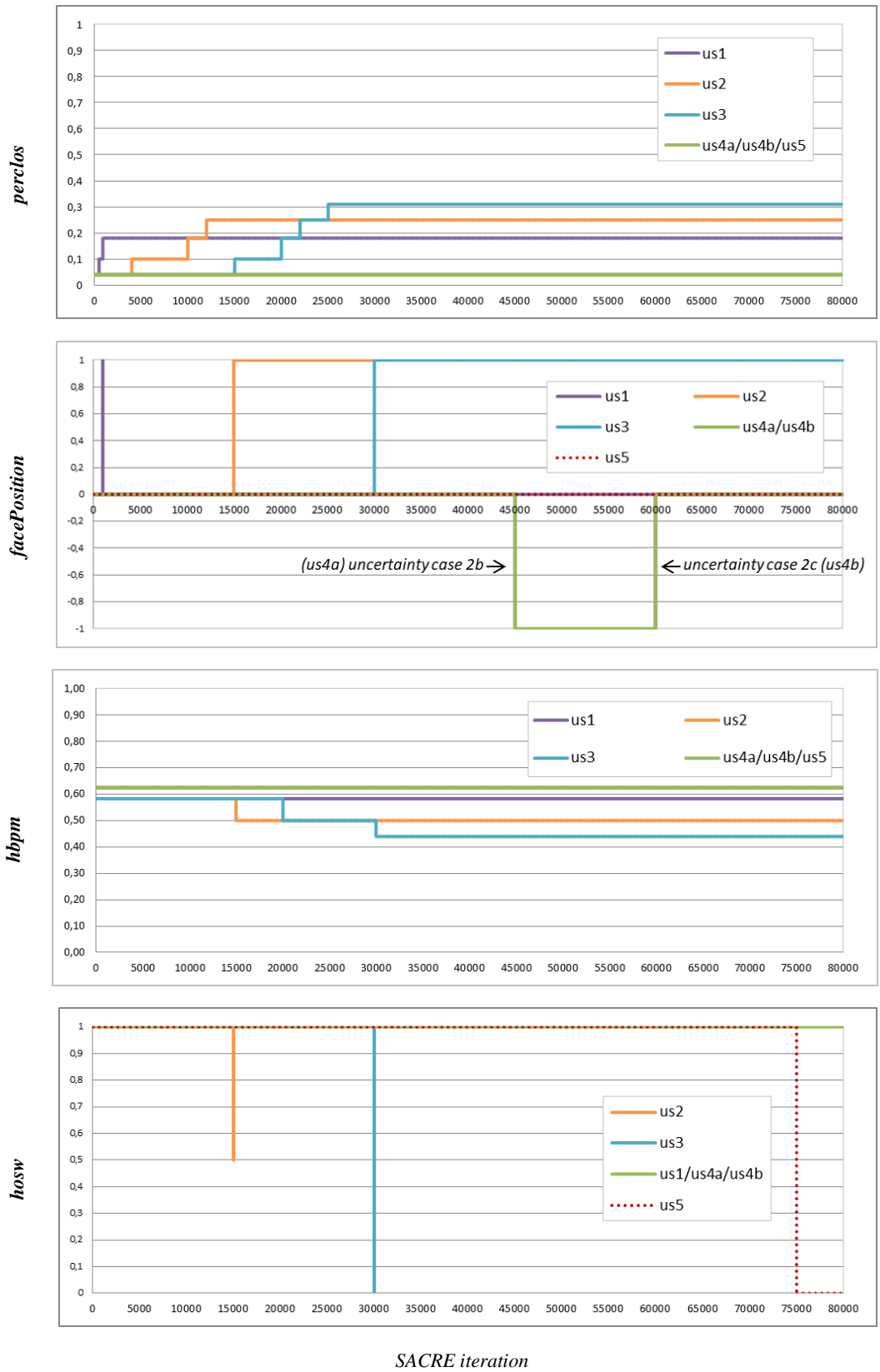

*Fig. 11. Sensors' variables values over execution time*

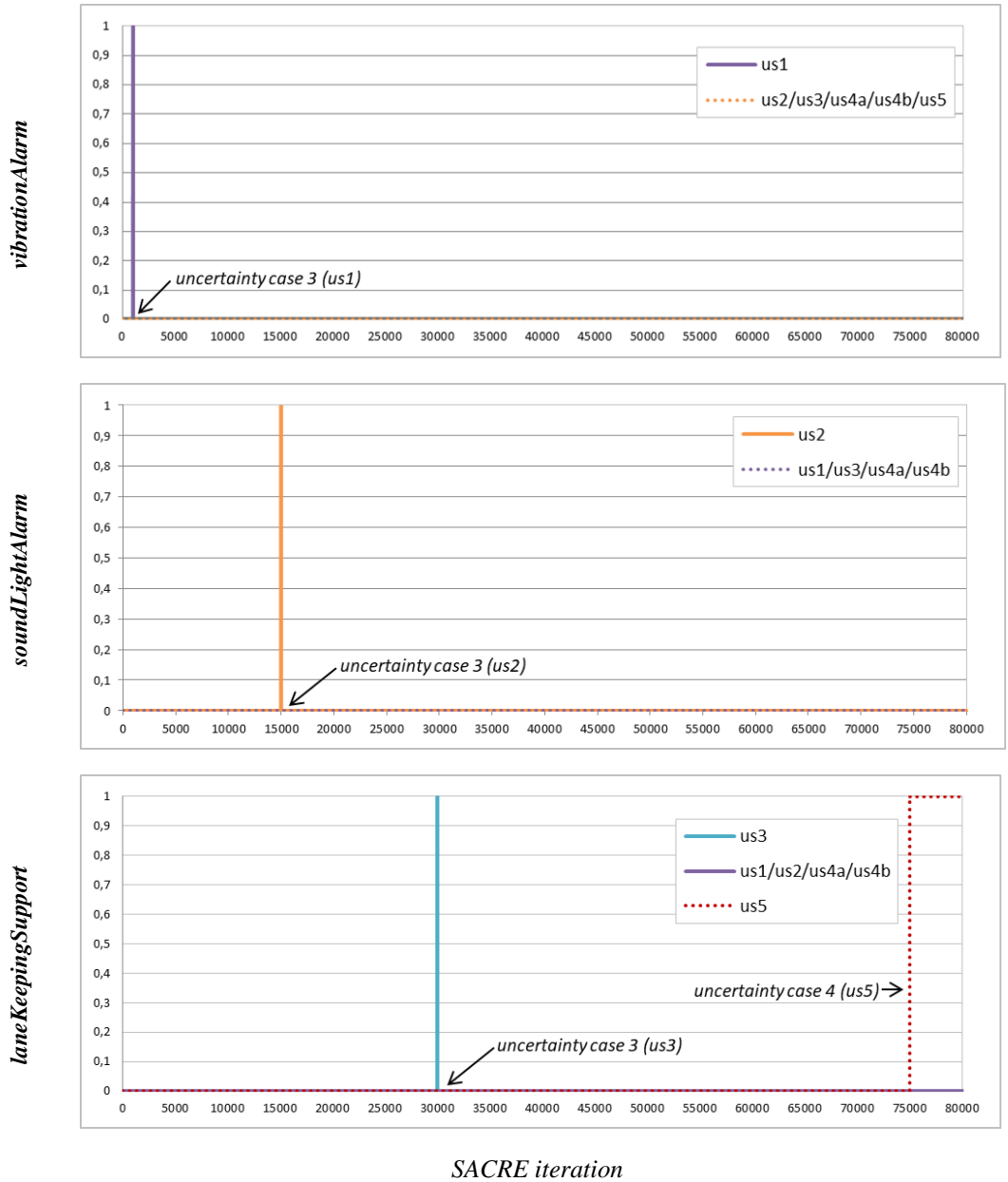

*Fig. 12. Actuators' state over execution time*

The resulting adapted contextual requirements' operationalization per scenario is presented in *Table* 7. We have also included the initial operationalization of the requirements' context (i.e., before the adaptation). In bold, we present the variables that have changed or have been added by SACRE in the adaptation process. In the cases of the variables that have been removed, we simply do not include them in the new operationalizations. Thus, in *Table* 7 we present the operationalizations as they were actually enacted in the smart vehicle. For **us$_5$**, two different valid adapted operationalizations resulted in the replications. In the rest of the scenarios, the same adapted operationalization resulted in all the replications.

Table 7. Resulting adapted contextual requirements

| Uncertainty scenario | CR adapted | Initial operationalization | Adapted normalized operationalization |
|---|---|---|---|
| us$_1$ | cr$_1$ | perclos>=0,15 AND hbpm<=0,60 AND hbpm >=0,56 | perclos>=0,15 AND hbpm<=0,60 AND hbpm>=0,56 **AND facePosition=1** |
| us$_2$ | cr$_2$ | perclos>=0,21 AND facePosition=1 AND hbpm<=0,55 AND hbpm>=0,46 | perclos>=0,21 AND facePosition=1 AND hbpm <=0,55 AND hbpm >=0,46 **AND hosw=0,5** |
| us$_3$ | cr$_3$ | perclos >0,30 AND facePosition=1 AND hbmp<=0,45 AND hosw <1 | perclos >0,30 AND facePosition=1 AND hbpm <=0,45 AND **hosw=0** |
| us$_{4a}$ | cr$_2$ | perclos>=0,21 AND facePosition=1 AND hbpm<=0,55 AND hbpm>=0,46 | perclos>=0,21 AND hbpm <=0,55 AND hbpm >=0,46 |
| us$_{4a}$ | cr$_3$ | perclos >0,30 AND facePosition=1 AND hbpm<=0,45 AND hosw <1 | perclos >0,30 AND hbpm <=0,45 AND hosw<1 |
| us$_{4b}$ | N/A | N/A (facePosition variable added to the active variables set for future operationalizations) | N/A (facePosition variable added to the active variables set for future operationalizations) |
| us$_5$ | cr$_3$ | perclos >0,30 AND facePosition=1 AND hbpm<=0,45 AND hosw <1 | perclos **<0,05** AND facePosition=**0** AND hbpm <=**0,75** AND **hbpm>=0,56** AND hosw<1 OR perclos **<0,05** AND face position=**0** AND hbpm <=**0,75** AND **hbpm>=0,56** AND hosw=**0** |

## 8.3. Analysis of the results

We analyze the results of the requirements' context operationalization adaptation 1) exploring SACRE's resulting response time (time elapsed since a Symptom is triggered until the adaptation is enacted) and 2) statistically analyzing the data mining algorithm.

- *Response time results:* In *Table 8* we provide the replications' average response time (in milliseconds) for each of the uncertainty scenarios executed in SACRE. Moreover, in order to evaluate the robustness of SACRE, we included the standard deviation of the response times.

Table 8. SACRE response time

| Uncertainty scenario | Adaptation response time (ms) | Adaptation response time standard deviation ($\sigma$) |
|---|---|---|
| us$_1$ | 3.859,50 | 1.004,60 |
| us$_2$ | 9.271,46 | 2.229,70 |
| us$_3$ | 13.358,25 | 3.028,38 |
| us$_{4a}$ | 2.477,52 | 689,39 |
| **us₄ᵦ** | 261,63 | 124,31 |
| **us₅** | 30.262,09 | 5402,64 |

Fig. 13 presents the detailed response time values obtained in each of the replications of the six uncertainty scenarios, represented each in a separate sub-graph. The *x*-axis of each sub-graph shows the number of replication (from 1 to 100), while the *y*-axis the adaptation response time.

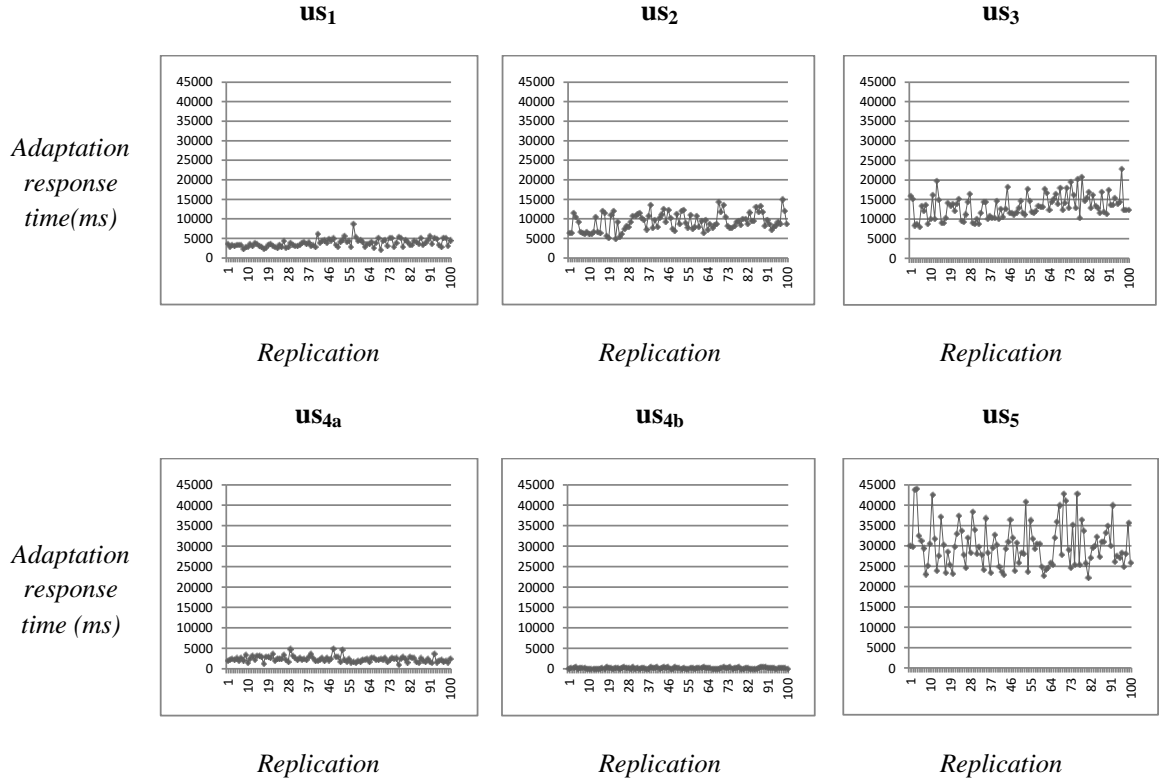

*Fig. 13. Adaptation response time per uncertainty scenario replication*

The response time values obtained in the different uncertainty scenarios going in average: from 3,85 sec. to 30,26 sec., for scenarios where data mining was required; and from 0,26 sec. to 2,47 sec., for scenarios that do not require data mining. For the first type of scenarios, graphs in Fig. 13 suggest a correlation between the amount of data to analyze (dictated by the SACRE iterations assigned in *Table 6*) and the experienced adaptation response time. Thus, we calculated the Pearson Product-Moment Correlation Coefficient (PPMCC) for the scenarios' SACRE iterations and the corresponding average response times. We obtained a coefficient of 0,99 which corroborates the existence of a correlation.

- *Statistical analysis:* Similarly to in our previous work (Knauss et al., 2016), we use a *10-fold cross validation* for statistically analyzing the data mining algorithm results. In SACRE the *10-fold cross validation* is executed at runtime every time the data algorithm is called. If the resulting *precision, recall and f-measure* are above the acceptance thresholds indicated in the Plan element, requirements' adaptation is accepted. We present in *Table* 9 the average resulting values of *precision, recall* and *f-measure* in each of the uncertainty scenarios. We have average the values reported in each scenario replication when an adaptation is accepted. Uncertainty scenarios **us$_{4a}$** and **us$_{4b}$** are not included in the table since they did not request the data mining algorithm.

*Table 9. SACRE data mining algorithm measures*

| Uncertainty scenario | Precision | Recall | f-measure |
|---|---|---|---|
| **us$_1$** | 1 | 1 | 1 |
| **us$_2$** | 1 | 1 | 1 |
| **us$_3$** | 1 | 1 | 1 |
| **us$_5$** | 0,969059 | 1 | 0,984252 |

Fig. 14 provide the details about the resulting data mining measures in each of the replications of the six uncertainty scenarios, represented each in separate sub-graphs. The *x*-axis of each sub-graph shows the number of replication (from 1 to 100), the *y*-axis the measure *precision*, *recall* or *f-measure* accordingly. As it can be noticed, the results of the measures were very high: invariantly 100%, for each of the measures in uncertainty scenarios **us$_1$**, **us$_2$** and **us$_3$**; and, 96,90%, 100% and 98,42%, for the precision, recall and f-measure respectively for the uncertainty scenario **us$_5$**. Variations in measures values, when existent, are presumably very small, thus we do not fourthly statically analyze them.

It is worth to mention that particularly, in the uncertainty scenario **us$_5$**, the data mining algorithm presented an output variation between replications, generating two different but still valid, operationalization's adaptation. The factors generating these variations could also explain the resulting measures in this uncertainty scenario. However, the study of the variation due to the internal operation of the data mining algorithm is out of the scope of this work. For better understanding of this and other data mining algorithm we refer the reader to previous works (Knauss et al., 2016; Rook, 2014; Rook et al., 2014).

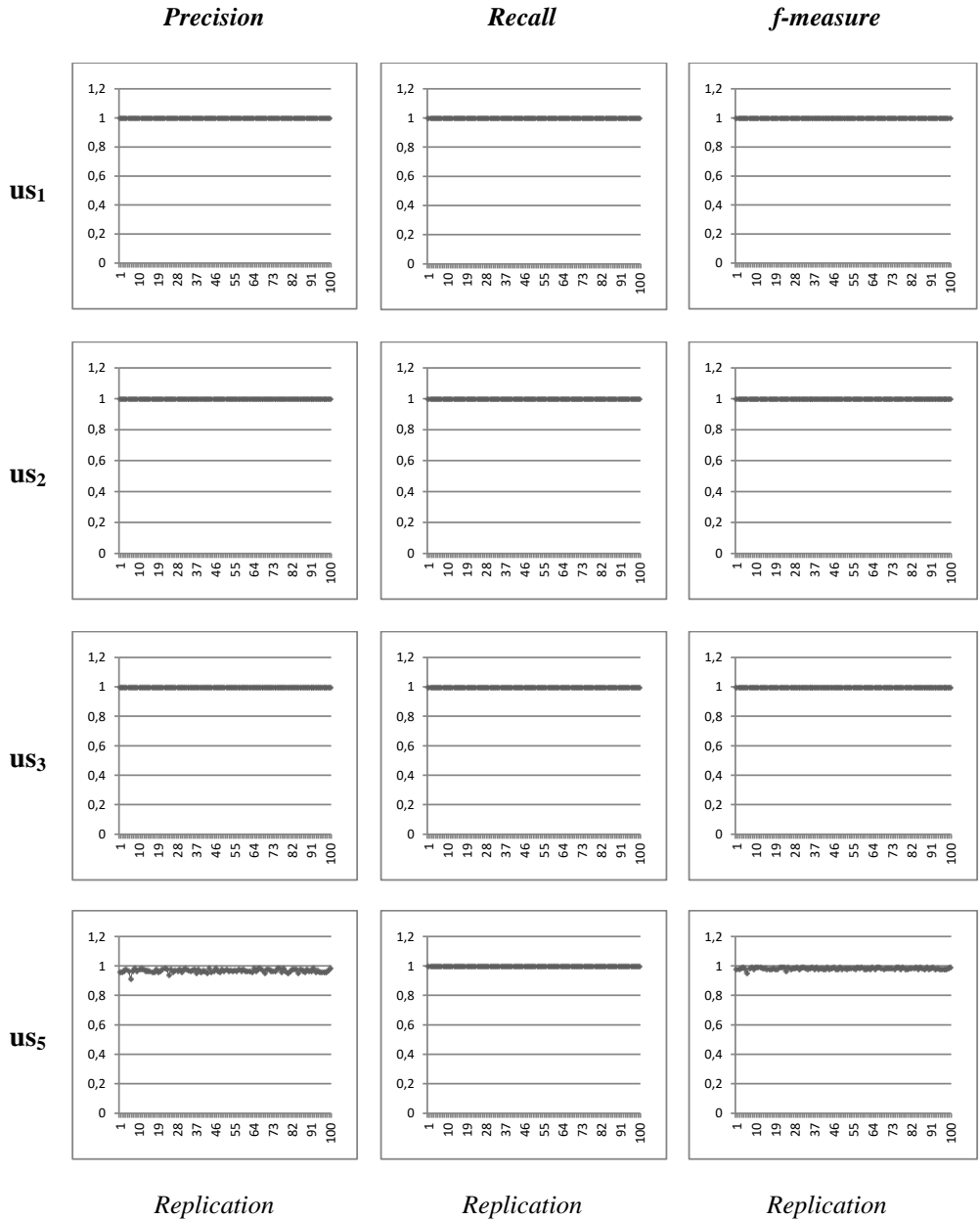

Fig. 14. Data mining algorithm measures per uncertainty scenario replication

## 8.4. Threats to validity

a) **Internal validity**. The internal validity of the evaluation concerns our ability to draw conclusions about the connections between the uncertainty scenarios and the resulting SACRE's adaptation response time and data mining measures. In order to reduce this kind of threat, we quantitatively interpret our results using descriptive statistics for determine tendencies, dispersion and dependencies.

b) **Construct validity**. A threat to construct validity is that our evaluation of SACRE is based on a simulated environment in which sensors' and actuators' variables values were designed by the experimenters. Thus, our evaluation could be affected by our interpretation of contexts and interactions of the driver with the smart vehicle. In order to reduce this threat, we study each of the variables to simulate in existing works of the domain and tried to model each of the variables as much close as possible to a real behavior, independently and in conjunction.

c) **Conclusion validity.** In our evaluation, we used six uncertainty scenarios for assessing SACRE. For the ones using data mining (us1, us2, us3 and us5), we achieved results of 100% for three out of four scenarios for the measurements precision, recall and f-measure. For the other scenario, results were over 96%. Although these excellent results regarding the data mining measures, the variations in the adaptation response times, particularly in this scenarios, was not so good. Variations in response time can be due to a diversity of factors, must of them not under our control. For instance, hardware resources available at a specific point, operating system executions management, Java ME emulator response time. Future work should investigate how these variation could be reduce, particularly in critical applications such the smart vehicles.

d) *External validity*. External validity refers to the generalizability of our conclusions. In this work, we have evaluated SACRE in the extremely demanding domain of smart vehicles. The results were satisfactory. However, due to the simulated environment in which the evaluation has been executed, generalization may be limited. No only to the domain but to its application in the domain. In the future, we plan to conduct more experiments in other execution environments and application domains to study the effectiveness and generalizability of our approach. We also encourage other researches to evaluate SACRE in the application domains of their interest, using other analysis algorithms and techniques. Particularly, the exploitation of its decentralization and distribution capabilities is an interesting direction for research.

## 9. Discussion

In this section we discuss our approach SACRE. The biggest novelty of the SACRE approach is the combination of a successful technique for supporting the adaptation of SASs contextual requirements in the presence of runtime uncertainty and extended well-known architectural styles for enabling the application of the adaptation technique in modern SASs. This allows SACRE to address four of the most important current open challenges affecting the engineering of SASs and their requirements (**Chl2.2**, **Chl2.3**, **Chl3.1** and **Chl3.2**). Although we have analyzed related work, other approaches with similar capabilities have not been found in literature. Thus, we consider SACRE an important advance in the fields of SASs and Requirements Engineering.

In this work, we have identified and listed the current open research challenges affecting the operation of SASs at runtime, then we reflected on related work regarding the most important and urgent challenges to address from the list. After that we present our approach SACRE as a solution for addressing the most urgent challenges with a single proposal and compare it with the related work. We have provided details about the construction of SACRE and supported SASs in the form of architectural and behavioral reference models. The models can be instantiated to fulfill a variety of system configurations and thanks to the knowledge distribution proposal and coordination mechanisms implementations in many domains are feasible. The full potential of SACRE will be discovered as new implementations and evaluations emerge.

In this work, an implementation of SACRE in the domain of smart vehicles is presented. We detail the whole process in a form of a guide for other developers and researchers. For the evaluation of SACRE, we executed different scenarios of uncertainty and demonstrate the feasibility of MAPE-K learning-based approaches, such SACRE, for supporting requirements' adaptation in the presence of runtime uncertainty in modern SASs, particularly in the smart vehicles domain. Statistical analysis results of the data mining algorithm were good in all the evaluated scenarios. Regarding the response time, we have found a strong correlation between the amounts of context data to analyze by the data mining algorithm and the adaptation response time of SACRE. Moreover, variations between executions of the same scenario in the adaptation response time were experienced. In the next section we provide the conclusion and discuss future work.

## 10. Conclusions and future work

In this paper we present SACRE, a step forward of our former approach ACon (Knauss et al., 2016). SACRE extends the behavioral proposal of ACon for supporting SASs contextual requirements' adaptation in the presence of runtime uncertainty, with a reference architecture and coordination mechanisms for applying the requirements' adaptation approach in modern, heterogeneous, decentralized and extremely demanding SASs. SACRE adopts the feedback control loop proposed by ACon for monitoring, analyzing and adapting contextual requirements. This loop relies on data mining techniques for determining best context operationalization of requirements affected by uncertainty at runtime. Unlike in our previous work (Knauss et al., 2016), in this paper we evaluate our approach, SACRE, through the implementation of the whole adaptation loop and evaluate its adaptations' response time and accuracy in the extremely demanding domain of smart vehicles.

The evaluation results demonstrate the feasibility and validity of a MAPE-K based approach such SACRE for supporting requirements' adaptation in modern SASs. Concretely, it corroborates its applicability and potential in the smart vehicles domain. In this evaluation we focused on contextual requirements for detecting and supporting drowsy drivers. In the future, we plan to implement SACRE in other execution scenarios for further demonstrating its validity, as well as in other application domains. We will also evaluate the impact of using other configurations of MAPE-K loops and re-configuring the elements at runtime (i.e., update configuration policies) in the adaptation response time and the data mining measures obtained in this work.

**Acknowledgments**

Thanks to CONACYT and I2T2, for the PhD scholarship granted to Edith Zavala. This work is partially supported by the SUPERSEDE project, funded by the European Union´s Information and Communication Technologies Programme (H2020) under grant agreement no. 644018 and partially funded by the Spanish project EOSSAC (TIN2013-44641-P).

# Appendix A. References identified in Stage 2

**Cheng et al., 2009**

**Weyns, 2017**

**Appendix B. References added in Stage 3**